\documentclass[twocolumn]{autart}    

\usepackage{graphicx}          
    \usepackage{epstopdf}                           
\usepackage{amsmath}
\usepackage{amssymb}  
\usepackage{cite}
\makeatletter
\let\NAT@parse\undefined
\makeatother

\usepackage{cases}
\usepackage{caption}
\usepackage{pgf, tikz}
\usepackage{cite}
\usepackage{fixmath}
\usepackage{mathptmx}
\usepackage{moresize}
\usepackage{graphicx}
\usepackage{epstopdf}

\usepackage{pgfplots}

\numberwithin{thm}{section}
 \numberwithin{assum}{section}
 \numberwithin{defn}{section}
 
 \numberwithin{rem}{section}
 
 \numberwithin{prop}{section}
 
 \numberwithin{lem}{section}
 \numberwithin{cor}{section}

\usetikzlibrary{shapes,arrows}
\usepackage{verbatim}

\DeclareSymbolFont{newfont}{OML}{cmm}{m}{it}
\DeclareMathSymbol{\Varrho}{3}{newfont}{37}
\usetikzlibrary{circuits.ee.IEC}

\usepackage{enumitem}
\DeclareMathAlphabet\mathbfcal{OMS}{cmsy}{b}{n}

\begin{document}

\begin{frontmatter}

\title{Granger Causality from Quantized Measurements\thanksref{footnoteinfo}} 

\thanks[footnoteinfo]{This work was partially supported by the Australian Research Council via Future Fellowship grant FT140100527.   The material in this paper was partially presented at the 58th IEEE Conf. Decision and Control (CDC), December, 2019, Nice, France \cite{CDC2019}  and  in  IFAC-PapersOnLine, after the recent cancellation of the 24th Int. Symp.  Mathematical Theory of Networks and Systems (MTNS 2020) \cite{MTNS2020}.   Corresponding author: Salman Ahmadi.}
\vspace{-5mm}
\author{Salman Ahmadi}\ead{ahmadis@student.unimelb.edu.au},    
\author{Girish N. Nair}\ead{gnair@unimelb.edu.au},               
\author{Erik Weyer}\ead{ewey@unimelb.edu.au}  
\address{Department of Electrical and Electronic Engineering, University of Melbourne, VIC 3010, Australia} 


\begin{abstract}                          
An approach is proposed for inferring Granger causality between jointly stationary, Gaussian signals from quantized data. First, a necessary and sufficient rank criterion for the equality of two conditional Gaussian distributions is proved. Assuming a partial finite-order Markov property,  a   characterization of Granger causality in terms of the rank of a matrix involving the covariances is presented. We call this the causality matrix. The smallest singular value of the causality matrix gives    a lower bound on
	the distance between the two conditional Gaussian distributions appearing in the definition of Granger
	causality and yields a new  measure of causality. Then, conditions are derived under which Granger causality between   jointly Gaussian processes   can be reliably inferred from the second order moments of  quantized measurements. A necessary and sufficient condition is proposed for Granger causality inference under binary quantization.  Furthermore, sufficient conditions are introduced to infer Granger causality between jointly Gaussian signals through measurements quantized via non-uniform,  uniform or high resolution quantizers. 
 Apart from the assumed partial Markov order and joint Gaussianity,  this  approach does not require  the parameters of a system model to be identified. No assumptions are made on the identifiability of the jointly Gaussian random processes through the quantized observations. The effectiveness of the proposed method is illustrated by simulation results. 
\end{abstract}

\end{frontmatter}

\section{Introduction}
Causal inference is the determination of the qualitative cause-and-effect (or
input versus output) relationships between two or more  signals over time. Sometimes these relationships are
obvious beforehand, but in many critical applications they are not. For instance, in environmental monitoring the direction in which a pollutant spreads may be unknown to begin with, making it difficult to determine {\em a priori} which measurements are inputs and which are outputs. In large manufacturing plants, the root cause of alarm signals is commonly obscured by complex feedback loops. A misunderstanding of the correct causal relationships not only reduces the accuracy of the subsequently
identified model, but could mislead decision-makers into poorly founded interventions.

In 1963, the econometrician C.~Granger introduced a definition of causality in terms of 
statistical prediction \cite{Granger1}, inspired by the work of N.~Wiener \cite{Wiener}. A signal $x$ is said to cause another signal $z$ if at some time, the optimal expected prediction error for a future value of $z$ is reduced by knowledge of past $x$ and $z$, as compared to if the past values only of $z$ are known.   In subsequent work \cite{GrangerNewbold1977, Granger3},   Granger proposed  a looser definition in terms of conditional probabilities, whereby $x$ is said to cause $z$ if, at some time, a future  $z$ and past $x$ are conditionally dependent given past  $z$; i.e., given past $z$, past $x$ can still influence the future of $z$.  In the case of jointly Gaussian processes under a mean-square error prediction error, the first definition coincides with the second. These definitions allow causality from $z$ to $x$ as well, which would reflect mutual coupling between the two processes as they evolve over time. 

As defined above, Granger causality is a `non-interventionist' notion based on signals rather than systems. This suits applications where  signals can only be measured, and cannot  easily be adjusted through experiments.  Under linear minimum mean-square error prediction, Granger \cite{Granger1969} and Sims \cite{Sims} showed the relationship between 	non-causality and zeroness of parameters  in  the vector autoregressive and moving average representations of the processes, respectively. 
In \cite{CainesChan}, a connection with linear systems theory is introduced through the idea of {\em feedback-freeness} for wide-sense stationary  vector   random processes  having a block triangular representation, which is shown to be equivalent to Granger non-causality under linear  minimum mean-square error prediction. Later in \cite{Caines}, a more restrictive version    called strong feedback free processes,  having block-diagonal innovation covariance  were  introduced  and their  equivalence to strictly causal linear systems  was  discussed. The relation between Granger non-causality and a linear time-invariant state space representation with a star graph structure as a network topology   was considered  in \cite{Jozsa}, and it was shown that Granger non-causality is equivalent to the existence of such a representation.   Granger causality between processes corrupted by additive noise or affected by filtering or sampling has also been investigated  \cite{Solo2007, Nalatore1, AndersonDeistlerDufour, Florin, Solo2016, IFAC2020}.

Based on the probabilistic definition of Granger causality, non-parametric measures of causality have been proposed in terms of {\em directed information} 
	\cite{Amblard, kontoyiannis2016estimating, quinn2011estimating, quinn2015directed}  and {\em transfer entropy} \cite{vicente2014efficient}. These measures are based on the Kullback-Leibler distance and take value zero iff there is no causality. Typically they also assume a finite joint Markov order for the joint process. For the special case of Gaussian processes, they generally reduce to the directed log-covariance measures introduced in \cite{geweke1982measurement}.

In this paper,  we focus on the effects of quantization    on Gaussian signals.  We investigate methods for inferring Granger causality
	between two jointly Gaussian, stationary signals using
	quantized measurements.  Quantization serves to reduce the communication load when transmitting sensor data.  In industrial systems, this facilitates efficient fault diagnosis and root-cause analysis when sensor readings exceed or drop below certain thresholds, without consuming excessive network bandwidth \cite{HuWang2017}.  Quantization is also potentially important in remote environmental monitoring, where a  strong causal relationship in one direction between two nearby field sensors may indicate the direction of spatial flow or movement of a population of animals, a pollutant, and so on.  Quantization here reduces transmission power and  prolongs battery life.  Evidently, the nonlinearity
	introduced by the quantizers moves this problem
	beyond the linear systems realm of the literature above.

Our main contributions are as follows:
\begin{itemize}
	\item  Assuming a {\em partial finite-order Markov}  property, 
	we introduce a {\em causality matrix} comprising joint process covariances, and show that Granger causality is characterized in terms of its rank. The basis of our analysis is a necessary and sufficient 	rank condition for two conditional Gaussian distributions to be identical (Theorem \ref{MainTheoremOfCGPDF}). This extends a recent 	result of \cite{Sullivant1, Sullivant2} on Gaussian conditional independence. To the best of our knowledge, this causality matrix has not been  previously studied in the system identification or causality literature. 
	\item  We present a geometric interpretation of the smallest singular value of the causality matrix which can be regarded as the     distance between the two conditional Gaussian distributions appearing in the definition of Granger causality, yielding a new  measure of the strength of causality. 
	\item We then consider binary quantization, employing  Van Vleck's formula \cite{VanVleckMiddleton} to express  the relation between the statistics of the quantized and unquantized signals. We  derive a necessary and sufficient condition to infer Granger causality between Gaussian signals from  binary measurements (Theorem \ref{CausalityForBinary}). This holds even though the variances of the unquantized signals may not be identifiable.
	\item Next we consider multi-level quantizers and derive sufficient conditions under which Granger causality between jointly Gaussian signals can be inferred from the second order moments of their quantized versions (Theorem \ref{GeneralTheoremForQuantization}). This uses a perturbation analysis of the causality matrix introduced above, bounds on covariances of quantization error terms combined with the Eckart-Young-Mirsky approximation theorem. Exploring this result, we derive sufficient conditions for three  cases: 
		\begin{enumerate}[label=\alph*)]
			\item non-uniformly quantized measurements with finite number of quantization levels (Proposition \ref{NonUniformTheorem}),
			\item  uniform quantization with infinitely many  quantization levels (Proposition \ref{MidTreadProposition}), 
			\item high resolution quantization (Proposition \ref{HighResProposition}).
		\end{enumerate}
 To express the relationships between the covariances of quantized and unquantized signals, we exploit Price's Theorem \cite{Price}  introduced in information theory and Widrow and Koll\'ar's results \cite{WidrowKollar}. 
\end{itemize}
Unlike much of the literature on causal inference, e.g. \cite{laghate2018learning, Sims, geweke1982measurement}, our approach does not require the statistics of the underlying Gaussian signals to be estimated, or a system model to be  explicitly identified apart from the finite partial Markov order. 

Preliminary versions of these results have been accepted or presented in \cite{CDC2019,MTNS2020}. Here, we provide full proofs for the binary and non-uniform quantization  cases, improving  the results compared with \cite{MTNS2020}, and refine  our previous results on high resolution quantization in \cite{CDC2019}.   We also introduce new results on uniform quantization with infinitely many levels, and  new results on the interpretation of the smallest singular value of the causality matrix as a measure of causality.   Moreover, we  investigate the required ergodicity properties of the   quantized data to be able to estimate their covariances.  Further, we present a numerical example.

 The  paper is organized as follows.   
 In Section \ref{GCInfWoQuan}, 
 we characterize  Granger causality between a pair of jointly Gaussian signals  in terms of the rank of a special matrix of covariances. Then we introduce an interpretation for the smallest singular value of the matrix as a measure of the strength of causality.  In  Section \ref{BinarySection} we derive necessary and sufficient conditions for Granger causality using binary quantized signals. In Section \ref{GCQuantized}, we investigate the effects of multi-level quantization and derive  sufficient conditions for inferring Granger causality between the unquantized signals from the statistics of the quantized data.  
  Using different techniques, we focus in Section \ref{GCUnderInfiniteLevel}  on infinite-level uniform quantization, including the high-resolution case.  The empirical estimation of second order moments of the quantized processes is discussed in  Section \ref{EmpiricalEstimation-sec}. Some simulation results are presented in Section \ref{Simulation-sec}  demonstrating the  proposed methods.  Section \ref{conclusions-sec} concludes the paper.

{\bf Notation:} Throughout this paper, we denote the random process segments  $(x_k)_{k=\ell}^n$ by $x^n_\ell$ , and $(x_k)_{k=1}^n$ by $x^n$. We use the conventions that  $x^n_\ell = (x_k)_{k=1}^n$ when $\ell\leq 1$, and 
equals the empty sequence when $\ell > n$ or $n<1$. Similarly, $x^n$ is the empty sequence when $n<1$. When clear from context, the full sequence $(x_k)_{k=1}^\infty$ is written as $x$. For a parameter $a$, the overline and underline ($\overline a$, $\underline a$) denote  known upper and lower bounds, respectively.  We also denote $\lim\limits_{y \to 0} \frac{g(y)}{f(y)} = 0$ by  the short-hand notation $f(y) \gg g(y)$.   




\section{Granger causality investigation} \label{GCInfWoQuan}

Let us first begin with the  definition of causality between discrete-time stochastic processes, in terms of conditional   independence. Let $x,z$ be discrete-time random  processes on the  time-axis $k=1,2,\ldots$ .   In 1980, Granger \cite{Granger3}  defined that  $x$ does not cause $z$ if: 
	\begin{align*} 
		\begin{split}
			P(z_{k+1}|x^k, z^k) = P(z_{k+1}|z^k), \ \  \\  \mbox{with probability (w.p.) }1, \ \ k=1,2,\ldots, 
		\end{split}
	\end{align*}
	where $P(\cdot|\cdot)$ denotes conditional probability measure.  Otherwise, if for some $k\geq 1$ there is a nonzero probability that $P(z_{k+1}|x^k, z^k) \neq P(z_{k+1}|z^k)$, then $x$ causes $z$. 

 In other words,  $x$ causes $z$ if and only if (iff)  there is a nonzero chance that at some time $k+1$, the value of $z$ could still be stochastically influenced by past $x$,  even if all past values of $z$ up to time $k$ are known.
If $x$ does not cause $z$,  the values of $z$ are always conditionally independent of past $x$, given past $z$.



Now let us assume the following:
 \begin{assum}[Partial Markov-$m$] \label{AssumptionOnOrderOFJoint} 
 The\hspace{.55mm}random process  $z$ is {\em partially Markov of order $m\geq 1$} in $x$ and $z$, that is, $P(z_{k+1}|x^k, z^k) =P (z_{k+1}|x^k_{k-m+1}, z^k_{k-m+1} ), \forall k \geq 1$. 
\end{assum} 

\begin{rem}
	 This is strictly weaker than being joint Markov for the same order $m$, and allows the $x$-process to potentially have a much higher-order dependence on the past. For instance, let us consider $x_{k+1} = a_0 x_k + a_1 x_{k-1} + b_0 z_k + b_1 z_{k-1} + w_k$ and $z_{k+1} = c_0 x_k  + d_0 z_k + v_k$ where $w_k$ and $v_k$ are independent white noise and $a_1$ or $b_1$ are nonzero. In this example, $x$ and $z$ are joint Markov of order two while the process $z$ is partially Markov of order one.  This example can easily be extended so that $x$ and $z$ are joint Markov of arbitrarily high or infinite order, while $z$ remains partial Markov-1.   In other words, the partial Markov assumption gives us extra flexibility. It can also be shown that having finite partial Markov orders in each direction does not necessarily imply a finite joint Markov order, unless $x_k$ and $z_k$ are also assumed to be conditionally independent given $x^{k-1}$ and $z^{k-1}$.    
\end{rem}

 With this restriction, we define Granger causality as follows:
\begin{defn}[Granger Causality]
	\label{GrangerCausality}
	The random process $x$ is said  {\em to not Granger cause (GC)}  $z$ if:
\begin{align} 
 P (z_{k+1}|x^k_{k-m+1}, z^k_{k-m+1} ) = P (z_{k+1}|z^k ),  \mbox{ w.p. 1}, \ \ \forall k\geq m. 
	\label{PracticalGCDef}
\end{align}
	  Otherwise, if for some $k\geq m$ there is a nonzero probability that $P (z_{k+1}|x^k_{k-m+1}, z^k_{k-m+1} ) \neq P(z_{k+1}|z^k)$, then $x$ is said to GC $z$.
\end{defn}
\begin{rem}
	We assess Granger causality definition from time $m$ rather than $1$ to avoid technical issues about initial conditions. Different authors have treated the initial condition differently; see e.g. \cite{FlorensMouchart}. In \cite{CainesChan, Solo2016, AndersonDeistlerDufour},  time starts from negative infinity  while Granger \cite{Granger3} considers the starting time $k=1$. In any case, it makes no practical difference because when we infer Granger causality, we need a large number of data points. 
	\end{rem} 
As the conditioning term on the RHS is no longer nested (i.e. included) in that of the LHS, this is no longer a conditional independence relationship.
In much of the literature e.g. \cite{kontoyiannis2016estimating, quinn2015directed, quinn2011estimating}, it is assumed that under non-causality the process $z$ is also Markov-$m$, so that $z^{k-m}$ can be dropped from the RHS. 
However, 
finite-order partial Markovianity does not generally imply finite-order marginal Markovianity.
In this paper, we do not make any {\em a priori} assumption on the Markovianity of $z$, but show that for jointly Gaussian signals, \eqref{PracticalGCDef} and Assumption \ref{AssumptionOnOrderOFJoint} imply that $z$ is marginally Markov-$m$, under a mild additional requirement.

For jointly Gaussian processes, it turns out that  \eqref{PracticalGCDef} can be expressed in terms of a rank condition. 
We first present this condition for general jointly Gaussian random vectors   $X,Y,Z,W$  with joint covariance matrix  $\Gamma$. The covariance matrices between subsets of variables are denoted by matching subscripts, for instance 
 $\Gamma_{X,Y} = \Gamma_{Y,X}^T$  is the cross-covariance matrix between  $X$  and  $Y$, while  $\Gamma_{[Z W], [Z W]}$ is the covariance matrix for the random vector  $[Z \ \ W]$.\footnote{For convenience we adopt this mild abuse of notation, rather than  $[Z^T W^T]^T$.}
\begin{thm} \label{MainTheoremOfCGPDF}
	Let   $X, Y,Z,W$  be jointly Gaussian random vectors with joint covariance matrix  $\Gamma$, and suppose that  the random vector  $[Y \ \ Z \ \ W]$  has positive definite covariance.
	Then the following statements are equivalent:
	\begin{enumerate}[label=\arabic*.]
	\item 		The conditional distributions $P( X|Z,Y )$ and $P( X|Z,W )$ are identical.
		\item
     	\begin{align} \label{MainEqualityOfCGPDF}
			\mathrm{rank}
			\left[
			\begin{array}{ccc}
				\Gamma_{ X,Y}   & \Gamma_{  X,Z } & \Gamma_{ X,W} \\
				\Gamma_{ Z,Y}  & \Gamma_{ Z,Z}  & \Gamma_{ Z,W} 
			\end{array}
			\right] =\#  Z,
		\end{align}
		where $\# Z$ is the dimension of the random vector   $Z$.
		\item   $P( X|Z,Y) = P( X|Z,W) =P( X|Z)$.   \hfill $\blacksquare$
	\end{enumerate}
\end{thm}
\begin{pf}
 See Appendix \ref{Proof_Of_MainTheoremOfCGPDF}. 
\end{pf}
\begin{rem}
	The second item in this result is a variation of a recent rank formula for Gaussian conditional independence, due to Sullivant \cite{Sullivant1, Sullivant2}. 
	Under joint Gaussianity, the positive definiteness of $\Gamma_{ [Y Z W], [Y Z W]}$ excludes degenerate cases where deterministic (affine) relationships exist between 
	 $Y$, $Z$ and  $W$.
	With additional analysis, it can be shown that if  $W$ is constant (requiring this positive definiteness  to be relaxed), then the formula of \cite{Sullivant1, Sullivant2} for conditional independence can be recovered as a special case.
	However, this extension is not needed for our purposes.  
	The third item in this result shows that $P( X|Z,Y ) = P( X|Z,W )$ if and only if conditional independence between  $X,Y$ and between  $X,W$  immediately follow given  $Z$. 
\end{rem} 
Now suppose that $x,z$ are  jointly stationary Gaussian signals satisfying  Assumption \ref{AssumptionOnOrderOFJoint}. Further assume that they are each scalar-valued, for simplicity, and that there is never a deterministic relationship between $x^k_{k-m+1}$ and $z^k$. 
  \begin{defn}[Causality Matrix]	  The $(m+1) \times (\ell+m)$ {\em causality matrix}  $C_G^{x \to z}(m,\ell), \ell \geq m$ is defined as: 
\begin{align} 
	{C_G^{x \to z}(m,\ell)} := 
	\left[
	\begin{array}{ccc}
		\Gamma_{z^*,\tilde{x}} & \Gamma_{z^*,\tilde{z}} &  \Gamma_{z^*, z^o}
	\end{array}
	\right],
	\label{CGdefeq}
\end{align}
where $z^* := z^{k+1}_{k-m+1}$, $\tilde{z} := z^k_{k-m+1}$,  $z^o := z^{k-m}_{k-\ell+1}$, and $ \tilde{x} := x^k_{k-m+1}$.   The covariances $\Gamma_{z^*,\tilde{x}}$ and $\Gamma_{z^*,\tilde{z}}$ are $(m+1) \times m$ matrices, and $\Gamma_{z^*,z^o}$ is an $(m+1) \times (\ell-m)$ matrix.
\end{defn} 

 The causality matrix depends on the cross-covariances between $x$ and $z$ and the autocovariances of $z$, but not on the autocovariances of $x$.  Let us first make the following assumption:
\begin{assum}\label{AssumptionCommon}
	Let $x,z$ be  jointly  stationary scalar Gaussian signals satisfying  Assumption \ref{AssumptionOnOrderOFJoint}. Further assume that there is no deterministic relationship between any of the components of $\left [ x^k_{k-m+1}  \ \ z^k\right ]$ at any time $k \geq m$. 
\end{assum} 
We have the following:
\begin{thm} \label{MainTheoremOfCG}  
Suppose Assumption \ref{AssumptionCommon} holds. 
\begin{enumerate} [label=\roman*)]
		\item $x$ does not Granger cause $z$ if and only if 
	\begin{align} \label{rank}
	 	\mathrm{rank}\hspace{.5mm} C_G^{x \to z}(m,k) = m 
			, \ \ \forall k \geq m.
		\end{align}
		\item If $x$ does not Granger cause $z$, then $z$ is marginally Markov-$m$, i.e.
		\begin{align} 
			P(z_{k+1}|z^k)=  P (z_{k+1}| z^k_{k-m+1} ).
		\end{align}
		\item If $P(z_{k+1}|x^k_{k-m+1}, z^k_{k-m+1})=  P (z_{k+1}| z^k_{k-m+1} )$ then $P(z_{k+1}|z^k)=  P (z_{k+1}| z^k_{k-m+1} )$. \hfill $\blacksquare$
\end{enumerate}
\end{thm} 
\begin{pf}
	 See Appendix \ref{Proof_Of_MainTheoremOfGNC}. 
\end{pf}
\begin{rem}
	A distinguishing feature of this result is that it allows Granger causality to be inferred directly from the second-order statistics of the signals, without having to analyze or fit a linear dynamical model as is often done in the literature.
\end{rem}

\begin{rem}
	The assumption of no deterministic relationship between any of the components of $\left [ x^k_{k-m+1} \ \ z^k\right ]$ in Theorem \ref{MainTheoremOfCG} is more relaxed than Axiom B in  Granger's paper \cite{Granger3} which assumes no deterministic relationship between any of the  components of $\left [ x^k\ \  z^k\right ]$. For instance, the process $x_{k+1} = x_{k-m+1}$ where $x_1,..., x_m$  are independent and identically distributed is covered by Theorem \ref{MainTheoremOfCG} but not by  \cite{Granger3}. This is a stationary random process which evolves periodically from its initial conditions. The last item in Theorem \ref{MainTheoremOfCG} does not require the assumption of no deterministic relationship between any of the component of  $\left [ x^k_{k-m+1} \ \ z^k\right ]$ at any time $k \geq m$.
\end{rem}
\begin{rem} \label{LastColumn}
	One of the implications of the last item in Theorem \ref{MainTheoremOfCG}  is that the last block column in \eqref{CGdefeq} is not needed when the rank condition in \eqref{rank} is evaluated, that is \eqref{rank} can be replaced by $\mathrm{rank} C_G^{x \to z}(m,m) = m$. However the last block column is kept  since it is used in the following Sections when sufficient conditions in terms of the smallest singular value are derived.
\end{rem}

 \subsection{Measure of causality}  \label{MeasureOfCausality} 
We now  introduce a geometric interpretation of the smallest singular value of the matrix introduced in Theorem \ref{MainTheoremOfCGPDF}. To do so, we consider the distance between the two conditional means of $P(X|Z,Y)$ and $P(X|Z,W)$
. The distance is measured by the variance of their differences. Under the assumption of joint Gaussianity, equality of the conditional means implies equality of the conditional distributions.  Hence this variance is a measure of distance between the conditional distributions. We show here that the smallest singular value of the matrix appearing in Theorem \ref{MainTheoremOfCGPDF} gives  a lower bound on this distance. 
	
	\begin{thm} \label{LowerBoundOnDifferenceBetweenConditionalMeans}
		Let $X, Y,Z,W$ be jointly Gaussian random vectors with joint covariance matrix  $\Gamma$, and suppose that  the random vector $[Y \ \ Z \ \ W]$ has positive definite covariance. Further assume that $X$ is scalar.\\
		The variance of difference between the conditional means of distributions $P(X|Z,Y)$ and $P(X|Z,W)$ is lower bounded as follows:
		\begin{align} \label{FINAL_1_Theorem_DifCondMeans}
			\text{Var} \Big({\mu_\text{cond}^{X|Z,Y}} - {\mu_\text{cond}^{X|Z,W}}\Big) \geq & \phi(\Gamma_{[YZW],[YZW]}) \times \nonumber\\
			&	\sigma_{\min}^2 \Bigg( \left[\begin{array}{ccc}
				\Gamma_{X,Y}  & \Gamma_{X,Z}  & \Gamma_{X,W} \\
				\Gamma_{Z,Y}  & \Gamma_{Z,Z}  & \Gamma_{Z,W} 
			\end{array}\right] \Bigg) ,
		\end{align}	
		where the positive
		\begin{align} \label{Defphi}
			\phi(\Gamma_{[YZW],[YZW]}) := &
			\min\bigg\{ \lambda_{\#Z+1}^2 \big( \Gamma_{[ZY],[ZY]}    \big),  \lambda_{\#Z+1}^2 \big( \Gamma_{[ZW],[ZW]}\big)  \bigg\} \times \nonumber \\ 
			& \lambda_{\min} \big( \Gamma_{[YZW],[YZW]} \big) , 
		\end{align}  mainly depends on the statistics of $Y,Z,W$ and does not depend on the random variable $X$.         \hfill$\blacksquare$
\end{thm} 
\begin{pf}
	See Appendix \ref{ProofOfGeometric}. 
\end{pf}
Obviously $\min\big\{ \lambda_{\#Z+1}^2 \big( \Gamma_{[ZY],[ZY]}    \big),  \lambda_{\#Z+1}^2 \big( \Gamma_{[ZW],[ZW]}\big)  \big\}$ in \eqref{Defphi} is always positive due to the assumption on the positive definiteness of the covariance matrix of the random vector $[Y \ \ Z \ \ W]$. Hence,  the only way to have the lower bound zero is:
\begin{align} \label{MatrixOfOurInterest}
	\sigma_{\min} \Bigg( \left[\begin{array}{ccc}
		\Gamma_{X,Y}  & \Gamma_{X,Z}  & \Gamma_{X,W} \\
		\Gamma_{Z,Y}  & \Gamma_{Z,Z}  & \Gamma_{Z,W} 
	\end{array}\right] \Bigg) = 0.
\end{align}

In this case, the matrix of interest is not full rank and Theorem \ref{MainTheoremOfCGPDF} implies that the conditional distributions coincide. As  the smallest singular value  increases, the lower bound on the distance between the two conditional distributions  grows.

 Based on the analysis above, the smallest singular value of the causality matrix $ C_G^{x \to z} (m,k) $   lower-bounds the distance between  the two conditional Gaussian probability measures appearing in the definition of Granger causality. We interpret  $\sigma_{\min}(C_G^{x \to z} (m,k))$  as a measure of the strength of   Granger causality.  Comparing with Geweke's measure of causality \cite{geweke1982measurement},  some of the properties of the measure $\sigma_{\min}(C_G^{x \to z} (m,k))$ introduced in this paper are as follows:
\begin{itemize}
	\item  $ \sigma_{\min} ( C_G^{x \to z} (m,k))  \geq 0$.
	\item   Like Geweke's measure, 	$\sigma_{\min} ( C_G^{x \to z} (m,k))=0 $   if and only if  $x$ does not Granger cause $z$.
	\item  The smallest singular value has a geometric interpretation as a lower bound on  the distance between the two conditional Gaussian probabilities. It is not scale-invariant while Geweke's measure is scale-invariant, being the ratio between two conditional variances.  
\end{itemize}


\section{Granger causality under binary quantization} \label{BinarySection}

 In this Section, we consider zero-mean jointly Gaussian stationary  signals $x$ and $z$ passing through binary quantizers with zero thresholds.  The zeroness of the mean can be checked by counting if the numbers of the values of the quantized outputs at both quantization levels are equal. Alternatively, a DC notch filter can be applied before quantization to enforce the zero mean. Note that preprocessing of data before quantization and transmission to the central data analysis point is standard, see e.g. \cite{Moffet1982}. Preprocessing occurs on signals $x$ and $z$ separately. Therefore the  Granger causality between $x$ and $z$ must be assessed at a central data analysis point, using the quantized versions of both.   If a mean is nonzero, this is equivalent to assuming that the respective quantizer threshold is set equal to it. The observed signals are $x^Q$ and $z^Q$, respectively.
We are interested in revealing whether or not $x$ Granger causes $z$, using the statistics of the quantized signals only.

In the causality matrix $C_G^{x \to z}(m,\ell)$ defined in \eqref{CGdefeq}, covariances ($\gamma_{xz}(.)$ and $\gamma_{zz}(.)$)  and the variance of signal $z$ ($\gamma_{z}^2$) are required. However, the variance of zero-mean Gaussian signals passed through a binary quantizer with threshold zero   cannot be estimated, e.g. \cite{MasryCambanis}. Hence, some of the elements of $C_G^{x \to z}(m,\ell)$ cannot be estimated. 

In order to solve this issue, a modified matrix is constructed. We know that if $B$ is a nonsingular matrix, then $\mathrm{rank} (C_G^{x \to z}(m,\ell) B) = \mathrm{rank} \hspace{.5mm} C_G^{x \to z}(m,\ell) $.  Let us construct a diagonal matrix $B := \text{diag} (\frac{1}{\gamma_{x}\gamma_{z}}, \dots, \frac{1}{\gamma_{x}\gamma_{z}}, \frac{1}{\gamma_{z}^2}, \dots, \frac{1}{\gamma_{z}^2})$ where   the numbers of $\frac{1}{\gamma_{x}\gamma_{z}}$ and $\frac{1}{\gamma_{z}^2}$ entries  equal $m$ and $\ell$, respectively.   Now instead of determining the rank of the causality matrix  $C_G^{x \to z}(m,\ell)$, the rank of
\begin{align} \label{DefinitionOfR}
	{R_G^{x \to z}}(m,\ell):= {C_G^{x \to z}}(m,\ell) B ,
\end{align}
whose elements are the cross-correlation coefficients between signals $x$ and $z$ and the auto-correlation coefficients of signals $z$ is obtained. The relation between auto-correlation coefficients of a zero-mean Gaussian signal passing through the binary quantizer with zero threshold and outputs -1 and +1 and the covariance of output of the binary quantizer can be estimated through  Van Vleck's formula \cite{VanVleckMiddleton}. 

Suppose that two zero-mean jointly Gaussian random variables $w_1$ and $w_2$ with correlation coefficient $\rho_{w_1 w_2} \neq 1$ are quantized by binary quantizers with zero thresholds and the outputs are $w_1^Q \in \{-1,+1\}$ and $w_2^Q \in \{-1,+1\}$, respectively:
 \begin{align}
		w_i^Q= \begin{cases}
		 	 -1 & \quad w_i<0\\
			+1 & \quad 	w_i>0
		\end{cases}, \ \ i=1,2.
\end{align} 
It can be shown that a relation similar to Van Vleck's formula holds between auto-/cross-correlation coefficients of the jointly Gaussian random variables and the covariances of the quantized variables as follows:
\begin{align} \label{GeneralizationOFVanVleck}
	\rho_{w_1 w_2} =\sin \big(\frac{\pi}{2} \gamma_{w_1^Q w_2^Q }\big).
\end{align}
 In fact, in order to investigate Granger causality in this Section, the elements of the causality matrix are modified to construct the  matrix ${R_G^{x \to z}}(m,\ell)$ depending  just on the correlation coefficients. The reason is that the variances of zero-mean Gaussian signals after binary quantization with the threshold zero are not identifiable. However, the rank of ${R_G^{x \to z}}(m,\ell)$  equals the rank of the causality matrix.  

Using Theorem \ref{MainTheoremOfCG} and Remark \ref{LastColumn},  the following necessary and sufficient condition on Granger causality between signals quantized by binary quantizers can be stated:
\begin{thm} \label{CausalityForBinary} 
 	Let $x,z$ satisfy Assumption \ref{AssumptionCommon} and be zero mean. 	Then $x$  does not Granger cause $z$ if and only if  the matrix $R_G^{x \to z}(m,m)$  \eqref{DefinitionOfR}  is not full rank. The elements of the matrix $R_G^{x \to z}(m,m)$ are auto- and cross-correlation coefficients associated with $C_G^{x \to z}(m,m)$ ($\rho_{xz}$ and $\rho_{zz}$) obtained as follows:
	\begin{align}
		\rho_{xz} (i) &= \sin \big(\frac{\pi}{2} \gamma_{x^Q z^Q }(i)\big), i=1-m, \dots, m-1, m,\\
		\rho_{zz} (j) &=\sin \big(\frac{\pi}{2} \gamma_{z^Q z^Q }(j)\big), j=1,2, \dots, m,\\
		\rho_{zz} (0) &=1,
	\end{align}
	where $\gamma_{x^Q z^Q }$ and $\gamma_{z^Q z^Q }$ are, respectively, cross- and auto-covariance estimates of one-bit measurements observed through binary quantizers with zero thresholds  and values -1 and +1. \hfill $\blacksquare$
\end{thm}
\begin{rem}
	Note that the necessity and sufficiency of the result follows from the fact that \eqref{GeneralizationOFVanVleck} provides us a closed form relationship between  the correlation coefficients of the unquantized signals and the covariance of their quantized versions. Thus, the correlation coefficients of zero-mean jointly Gaussian signals can be obtained. Furthermore, the rank of the causality matrix and of the matrix constructed by correlation coefficients are equal. Hence, Theorem \ref{MainTheoremOfCG} and Remark \ref{LastColumn} imply the necessity and sufficiency on Granger causality between jointly zero-mean Gaussian, stationary random processes.
\end{rem}
\section{Inferring Granger causality using finite-level quantized data}
\label{GCQuantized}
In this Section,  the impact of finite-level quantization on the inference of Granger causality between the jointly Gaussian, stationary processes $x$ and $z$ where $z$ is partially Markov of order $m$ is investigated. Using the second-order statistics of the quantized data, we construct a  post-quantization matrix ${C^{x^Q \to z^Q}}$ that mirrors the causality matrix $C_G^{x \to z}$ \eqref{CGdefeq} of the unquantized processes.
We then show that if the difference between these two matrices is sufficiently small, then the full-rankness of  ${C^{x^Q \to z^Q}}$  implies that $x$ Granger causes (GC) $z$.
In the following Subsection,  some preliminary results useful for inferring Granger causality through quantized signals are presented.
\subsection{Preliminaries}


Theorem \ref{MainTheoremOfCG} implies that if  for some $q\geq m$ the causality matrix ${C_G^{x \to z}(m,q)}$  has full rank $m+1$, then   $x$ GC $z$.
The question here is whether we can infer  this causal relationship from the covariances of the quantized data. To answer this question,  we will use a classical result in linear algebra: 
\begin{thm} \textit{(Eckart-Young-Mirsky Matrix Approximation \cite{EckartYoung, Mirsky})}
	\label{EYM}
	Let the matrix $M\in \mathbb{R}^{l \times s}$ have rank $r$ and singular value decomposition   $M=\sum_{i=1}^{r}\sigma_i u_iv_i^T$, where $u_i,v_j$, $1\leq i,j\leq r$ are orthonormal vectors and  $\sigma_1 \geq \sigma_2 \geq \dots \geq \sigma_r (> 0)$ are the singular values.\\	
	If $p<r$, then
	\begin{align} \label{Mirsky}
		\min_{\mathrm{rank} \hspace{.5mm}X = p} \|M-X\|_2 = \| M- M_p \|_2= \sigma_{p+1},
	\end{align} 
	and
	\begin{align} \label{EckartandYoung}
		\min_{\mathrm{rank} \hspace{.5mm} X = p} \|M-X\|_F  = \| M- M_p \|_F= \sqrt{\sum_{i \geq p+1}\sigma_{i}^2}, 
	\end{align} 
	where $M_p=\sum_{i=1}^{p}\sigma_i u_iv_i^T$. \hfill $\blacksquare$
\end{thm}
\begin{rem}
	Unless otherwise stated,  $\|\cdot\|$ denotes either two-norm  ($\|\cdot\|_2$) or Frobenius norm ($\|\cdot\|_F$) in the following.
\end{rem}

\subsection{Granger causality inference under quantization} \label{GCIUQ}
In this Subsection, the  relationship between the causality matrix and its counterpart constructed by quantized signals is investigated. Using the results above, a	condition to determine the rank of the causality matrix through the rank of its counterpart ${C^{x^Q \to z^Q}}$ is derived.
\subsubsection{Relationship between ${C_G^{x \to z}}$ and ${C^{x^Q \to z^Q}}$} 
\label{QVsUnQ}
We begin with the relationship between the scalar covariances $\gamma_{z x}$, $\gamma_{z z}$ of the unquantized signals, and $\gamma_{z^Q x^Q}$, $\gamma_{z^Q z^Q}$ of their quantized versions, where superscript $^Q$ on a signal denotes the quantized version. For convenience, in the following we suppress time lags and use $w_1$ and $w_2$ to denote $x$ and/or $z$.
We have 
\begin{align} \label{RelationBetQandUnq}
	\gamma_{w_1^Q w_2^Q}= \gamma_{w_1 w_2} +\gamma_{w_1 \epsilon_2}+\gamma_{\epsilon_1 w_2}+\gamma_{\epsilon_1 \epsilon_2},
\end{align}
where $\epsilon_1$ and $\epsilon_2$ denote the quantization errors of $w_1$ and $w_2$ respectively  ($\epsilon_i := w_i^Q- w_i$). 
Now define the matrix
\begin{align} \label{CQDefinition}
	{C^{x^Q \to z^Q }} (m,\ell) :=& \left[
	\begin{array}{ccc}
		\Gamma_{{z^*}^Q,\tilde{x}^Q} & \Gamma_{{z^*}^Q,\tilde{z}^Q} &  \Gamma_{{z^*}^Q, {z^o}^Q}
	\end{array}
	\right] \nonumber \\
	\equiv& {C_G^{x \to z}} (m,\ell)  + {\Gamma_{\epsilon}}(m,\ell),
\end{align}
where $	{\Gamma_{\epsilon}}(m,\ell) 
:= [\gamma_{\epsilon_{ij}}] $ is defined as:
\begin{align} \label{RelatedToQuantizationNoiseCov}
	{\Gamma_{\epsilon}}(m,\ell) 
	:= &\left[
	\begin{array}{ccc}
		\Gamma_{z^*,\epsilon_{\tilde{x}}} & \Gamma_{z^*,\epsilon_{\tilde{z}}} &  \Gamma_{z^*, \epsilon_{z^o} }
	\end{array}
	\right]+ \left[
	\begin{array}{ccc}
		\Gamma_{\epsilon_{z^*}, {\tilde{x}}} & \Gamma_{\epsilon_{z^*} ,{\tilde{z}}} &  \Gamma_{\epsilon_{z^*}, z^o}
	\end{array}
	\right]+ \nonumber \\
	& \left[
	\begin{array}{ccc}
		\Gamma_{\epsilon_{z^*} ,\epsilon_{\tilde{x}}} & \Gamma_{\epsilon_{z^*}, \epsilon_{\tilde{z}}} &  \Gamma_{\epsilon_{z^*}, \epsilon_{z^o}}
	\end{array}
	\right], 
\end{align} 
which is the matrix version of the scalar relationship \eqref{RelationBetQandUnq}.
\begin{thm} \label{GeneralTheoremForQuantization} 
 
	Suppose Assumption \ref{AssumptionCommon} holds. 
	If there exists some $q\in [m,k]$ such that  the matrix ${C^{x^Q \to z^Q }}(m,q)$ in \eqref{CQDefinition}  involving the covariances of quantized data is full-rank with smallest singular value 
	\begin{align} \label{GeneralEquationForQuantization}
		\sigma_{\text{min}}({C^{x^Q \to z^Q}}(m,q)) > \| {\Gamma_{\epsilon}} (m,q)\|_2,
	\end{align}
	where $\Gamma_{\epsilon} (m,q)$ is defined in \eqref{RelatedToQuantizationNoiseCov}, then the unquantized Gaussian signal $x$ Granger causes the unquantized Gaussian signal $z$. \hfill $\blacksquare$
\end{thm}
\begin{pf}
	Denote the full-rank matrix ${C^{x^Q \to z^Q}}(m,q) $ by  $M$. If $X$ is any other matrix of the same dimensions but lower rank,
	then Eckart-Young-Mirsky matrix approximation Theorem (Theorem \ref{EYM}) states that  $\|X-M\|\geq \sigma_{\text{min}}(M)$. Conversely, if $\|X-M\|< \sigma_{\text{min}}(M)$, then
	$X$ remains full-rank. Setting $X ={C^{x \to z}}(m,q)  \equiv {C^{x^Q\to z^Q}}(m,q)  + {\Gamma_{\epsilon}}(m,q)$,
	we see that if $\sigma_{\text{min}}({C^{x^Q \to z^Q}}(m,q) ) > \| {\Gamma_{\epsilon}}(m,q)  \|$, then ${C_G^{x \to z}}(m,q)$ is guaranteed to be full-rank.
Theorem \ref{MainTheoremOfCG} then implies that $x$ GC $z$. 	
\end{pf}
\begin{rem} \label{Remark_One_For_GeneralTheoremForQuantization}
	This result states that GC can be inferred from the statistics of the quantized data, provided that the quantization perturbation, as measured by  
	$\| {\Gamma_{\epsilon}}(m,q)  \|$, is smaller than $\sigma_{\text{min}}({C^{x^Q \to z^Q}}(m,q) )$, which can be taken as a measure of how far ${C^{x^Q \to z^Q}}(m,q)$ is from losing full rank.
	However, both sides of  inequality \eqref{GeneralEquationForQuantization} depend on the quantization schemes. We can derive a condition that compares
	the size of the quantization perturbation to the statistics of the unquantized processes, as follows.
	We know that for two matrices $A$ and $B$ of the same dimension $ \sigma_{\text{min}}(A) - \sigma_{\text{max}}(B) \leq \sigma_{\text{min}}(A+B)$, (\hspace{-.4mm}\cite{horn2013matrix}, Problem 7.3.P16). Therefore, we have: 
	\begin{align} \label{Achievability1}
		\sigma_{\text{min}}({C_G^{x \to z}}(m,q)) - \sigma_{\text{max}}(\Gamma_{\epsilon}(m,q)) \leq \sigma_{\text{min}}({C^{x^Q \to z^Q}}(m,q)),
	\end{align}
	Furthermore, we know that $\|{\Gamma_{\epsilon}} (m,q)\|_2 =  \sigma_{\text{max}}(\Gamma_{\epsilon}(m,q))$. 
	So if
	\begin{align} \label{Achievability2}
		\| {\Gamma_{\epsilon}} (m,q)\|_2 < \sigma_{\text{min}}({C_G^{x \to z}}(m,q)) - \sigma_{\text{max}}(\Gamma_{\epsilon}(m,q)),
	\end{align}
	then by \eqref{Achievability1} it is guaranteed to be smaller than $\sigma_{\text{min}}({C^{x^Q \to z^Q}} \allowbreak (m,q) )$ as required.
	Finally, note that since $\|{\Gamma_{\epsilon}} (m,q)\|_2 =  \sigma_{\text{max}}(\Gamma_{\epsilon}(m,q))$, 
	the condition \eqref{Achievability2} may be equivalently written as
	\begin{align} \label{Achievability3}
		2\| {\Gamma_{\epsilon}} (m,q)\|_2 < \sigma_{\text{min}}({C_G^{x \to z}}(m,q)),
	\end{align}
	which can be achieved with sufficiently fine quantization iff $C_G^{x \to z}(m,q)$ is full-rank.
\end{rem}
\begin{rem}  
	Note that Theorem \ref{GeneralTheoremForQuantization} and Remark \ref{Remark_One_For_GeneralTheoremForQuantization}  hold not only for quantization but also 	for perturbations such as other nonlinearities or additive noise. In such cases, $\Gamma_{\epsilon} (m,q)$ is  the additive perturbation on the causality matrix and   the smallest singular value in  Theorem \ref{GeneralTheoremForQuantization}  corresponds to the matrix of covariances formed by the perturbed signals. 
\end{rem}
\subsection{Granger causality under non-uniform quantization}\label{NonuniformSection}


In this Subsection we consider a general case of Granger causal inference with non-uniform, finite-level quantization. With more than two levels, we no longer have recourse to \eqref{GeneralizationOFVanVleck}, which recovers the correlation coefficients of the unquantized signals from the covariances of their binary versions. 
Our approach here, as presented in Subsection \ref{GCIUQ}, is to formally construct and analyze a matrix of the same form as the causality matrix \eqref{CGdefeq}, but in terms of the covariances of the quantized processes. As  the quantized signals are not jointly Gaussian anymore, this matrix cannot be regarded as a causality matrix. We denote such a matrix by $C^{x^Q \to z^Q}(m,\ell)$. Using the perturbations caused by quantization on the covariances of unquantized signals and the rank of the matrix $C^{x^Q \to z^Q}(m,\ell)$, we then derive sufficient conditions to determine whether the causality matrix $C_G^{x \to z}(m,\ell)$ is full rank. Then Theorem \ref{MainTheoremOfCG} implies that $x$ Granger causes $z$  (Proposition \ref{NonUniformTheorem}).

In the following Subsection, the relation between covariances of quantized and unquantized signals is first derived, using Price's theorem \cite{Price, Papoulis1965}. Price's theorem has been previously used in the literature to study the determination   of autocorrelations and variances from data passed through quantizers or other nonlinearities \cite{CambanisMasry, MasryCambanis}.


\subsubsection{Relation between covariance of quantized and \- unquantized signals} \label{CovQuanVSUnquan}
 For convenience, in the following jointly Gaussian random variables are denoted by $w_1$ and $w_2$ which can be $x$ and/or $z$ at different times and lags.

Let us consider  two zero-mean jointly Gaussian random variables $w_1$ and $w_2$ with correlation coefficient $\rho \neq \pm 1$ that are quantized by non-uniform finite-level quantizers. Their quantized versions are denoted by $w_1^Q$ and $w_2^Q$, respectively:
\begin{align} 
	w_1^Q&=Q_1\big(w_1\big)=\sum_{i=1}^{n_1}l_{i}I_{c_{i-1}<w_1\leq c_{i}}, \label{QNon1}\\
	w_2^Q&=Q_2\big(w_2\big)=\sum_{j=1}^{n_2}l'_{j}I_{d_{j-1}<w_2\leq d_{j}},\label{QNon2}
\end{align}
where $n_1$ and $n_2$ are the number of quantization levels, $c_i$'s and $d_j$'s denote the thresholds of the quantizers and $l_{i}$ and $l'_{j}$ are levels of quantizers. Note that 
		 $c_0,d_0,c_{n_1}$ and $d_{n_2}$ can be infinite. Without loss of generality, suppose that the quantizer levels are increasing ($l_{i+1}> l_{i}, i=1,2, \dots, n_1-1$ and $l'_{j+1} > l'_{j}$, $j=1,2, \dots, n_2-1$). Recall that the joint probability density function of $w_1$ and $w_2$ is as follows:
\begin{align}
	f(w_1,w_2)&=\frac{1}{2\pi\gamma_{w_1} \gamma_{w_2}\sqrt{1-\rho^2}} e^{\frac{-1}{2(1-\rho^2)}\big(\frac{w_1^2}{\gamma_{w_1}^2}-\frac{ 2 \rho  w_1 w_2}{\gamma_{w_1} \gamma_{w_2}}+\frac{w_2^2}{\gamma_{w_2}^2}\big)},
\end{align}
where $\gamma_{w_1}$ and $\gamma_{w_2}$ are the standard deviations of the random variables $w_1$ and $w_2$, respectively.

First let us recall Price's theorem \cite{Price, Papoulis1965}. 
\begin{thm} [Price's Theorem]
	Let $w_1$ and $w_2$ be jointly Gaussian random variables with joint probability density function $f(w_1,w_2)$ and $g(w_1,w_2)$ is some function such that  $|g(w_1,w_2)| < A e^{|w_1|^\alpha + |w_2|^\alpha }$  where $ A>0, \alpha <2$ then
	\begin{align}
		\frac{\partial^n E\Big\{ g\big(w_1,w_2\big) \Big \}}{\partial \gamma_{w_1w_2}^n} &= E\Big\{\frac{\partial^{2n} g(w_1,w_2)}{\partial w_1^n \partial w_2^n}\Big\}, \label{Price0}
	\end{align}
	where $\gamma_{w_1w_2}$ is covariance between $w_1$ and $w_2$.\hfill $\blacksquare$
\end{thm}
Using Price's Theorem, the covariance between quantized signals $w_1^Q$ and $w_2^Q$ can be obtained as follows.  See Appendix \ref{Derivation_Of_Eq_CovNonUniformMultilevel1} for derivation of \eqref{CovNonUniformMultilevel1}: 
\begin{align} \label{CovNonUniformMultilevel1}
	\gamma_{w_1^Q w_2^Q } =&\sum_{i=1}^{n_1-1}\sum_{j=1}^{n_2-1}(l_{i+1}-l_{i})(l'_{j+1}-l'_{j}) \times \nonumber\\
	& \int\limits_{0}^{\rho} \frac{1}{2\pi\sqrt{1-y^2}} e^{\frac{-1}{2(1-y^2)}\big(\frac{c_i^2}{\gamma_{w_1}^2}-2 y \frac{c_i d_j}{\gamma_{w_1} \gamma_{w_2}}+\frac{d_j^2}{\gamma_{w_2}^2}\big)}  dy.  
\end{align}
Note that after quantization we do not have access to the values of correlation coefficients $\rho$ and of the standard deviations $\gamma_{w_1}$ and $\gamma_{w_2}$ (with one single subscript). 
In the following, we  find upper bounds on $|\gamma_{w_1^Q w_2^Q }-\gamma_{w_1 w_2 }|$ and then we exploit such bounds to guarantee the causality matrix is full rank. 
	In other words, to apply Theorem \ref{GeneralTheoremForQuantization} for non-uniform quantized signals, the norm of $	{\Gamma_{\epsilon}}(m,\ell) 
	= [\gamma_{\epsilon_{ij}}] $, whose components are the differences between covariances of unquantized and quantized signals $x$ and $z$ as defined in \eqref{RelatedToQuantizationNoiseCov}, needs to be bounded.

\subsubsection{Granger causality investigation through non-uniformly quantized signals}\label{ResultForNonUniform}
Let zero-mean jointly Gaussian stationary random processes $x$ and $z$ be quantized by non-uniform quantizers of the form \eqref{QNon1} and \eqref{QNon2}  with $n_x$ and $n_z$ levels, thresholds $c^x_i, i=1,...,n_x$  and $c^z_j, j=1,...,n_z$, and quantization levels  $l_{i}^x$ and $l_{j}^z$, respectively.   
For convenience, we assume that the quantizer levels are increasing. We wish to determine whether $x$ Granger causes $z$, using the statistics of the  quantized data $x^Q$ and $z^Q$.\\
 Our approach is to use upper bounds on $	|\gamma_{x^Qz^Q}(\kappa) - \gamma_{xz}(\kappa)| $ and $|\gamma_{z^Qz^Q}(\kappa) - \gamma_{zz}(\kappa)|$ to find an upper bound on the norm of ${\Gamma_{\epsilon}}(m,\ell)$. If this bound is less than $\sigma_{\text{min}}({C^{x^Q \to z^Q}}(m,q))$, then Theorem \ref{GeneralTheoremForQuantization} implies $x$ Granger causes $z$. Let 
\begin{align} 
	S_{xz}&:=  \max_{\substack{|\rho| \leq  \overline{\rho}_{xz} \\
			\underline \gamma_x   \leq \gamma_x\leq \overline \gamma_x   \\ \underline \gamma_z   \leq \gamma_z\leq \overline  \gamma_z }} |\gamma_{x^Qz^Q} - \rho \gamma_x \gamma_z|, \label{Easy_Sxz} \\
	S_{zz}&:=  \max_{\substack{|\rho| \leq  \overline{\rho}_{zz} \\
			\underline \gamma_z   \leq \gamma_z\leq \overline \gamma_z }} |\gamma_{z^Qz^Q} - \rho  \gamma_z^2| \label{Easy_Szz},\\
	S_{z}&:=  \max_{\substack{ \underline \gamma_z   \leq \gamma_z\leq \overline \gamma_z }} |\gamma_{z^Q }^2 -  \gamma_z^2| \label{Easy_Sz},	
\end{align} 
where $\gamma_{x^Qz^Q}$ and $\gamma_{z^Qz^Q} $ are functions of correlation coefficient $\rho$ given by \eqref{CovNonUniformMultilevel1} and  $\gamma_{z^Q }^2$ is given by $\int_{-\infty}^{+\infty} (Q(z)  - E\{Q(z)\})^2 f(z) dz$ where $f(z)$ is the univariate Gaussian density function. The parameters $\overline{\rho}_{xz}$ and $\overline{\rho}_{zz}$ are upper bounds on $ \max_{\kappa }|\rho_{xz}(\kappa)|$ and $  \max_{\kappa \neq 0}|\rho_{zz}(\kappa)|$ respectively, $\overline{\gamma}_z $ and $\overline{\gamma}_x $ are upper bounds on the standard deviations of $z$ and $x$, respectively, and  $\underline  \gamma_z  $ and $\underline  \gamma_x$ are, respectively, lower bounds on  the standard deviations of $z$ and $x$.  

 An upper bound  on $\| {\Gamma_{\epsilon}} (m,q) \|_2$ in \eqref{GeneralEquationForQuantization} is obtained as follows. We know that for $A = [a_{ij}] \in \mathbb{R}^{s \times t}$, we have $\|A\|_2 \leq \sqrt{  \|A\|_1  \|A\|_\infty}$  where  $ \|A\|_1 := \max_{1\leq j\leq t} \allowbreak \sum_{i=1}^s |a_{ij}|$ and $ \|A\|_\infty := \max_{1\leq i\leq s} \allowbreak \sum_{j=1}^t |a_{ij}|$.

Note that the elements of the causality matrix $C_G^{x \to z}(m,\ell)$ are  the covariances between $x$ and $z$ and the variance of and the covariances between $z$. The maximum difference between the components of the causality matrix $C_G^{x \to z}(m,\ell)$ and the matrix $C^{x^Q \to z^Q}(m,\ell)$ can be obtained using \eqref{Easy_Sxz}-\eqref{Easy_Sz}   or through Appendix \ref{PerturbationBound}. First upper bounds on $\| {\Gamma_{\epsilon}} (m,q) \|_1$ and  $\| {\Gamma_{\epsilon}} (m,q) \|_\infty$ are derived as follows:
	\begin{align} \label{BoundForOneNorm_NonUniform}
		\|{\Gamma_{\epsilon}} (m,q) \|_1 \leq  N_o := \max \big\{  (m+1) S_{xz}, (m+1) S_{zz},           m S_{zz}  +  S_{z}     \big \},
	\end{align}	
and
\begin{align} \label{BoundForInfty_NonUniform}
	\| {\Gamma_{\epsilon}} (m,q) \|_\infty \leq N_I:=& m S_{xz} +(q-1) S_{zz}  +\max \big\{  S_{zz},      S_{z}   \big \}.
\end{align}	
Using  \eqref{BoundForOneNorm_NonUniform},  \eqref{BoundForInfty_NonUniform} and Theorem \ref{GeneralTheoremForQuantization}, the following can be presented:
\begin{prop} 
	\label{NonUniformTheorem}
 	Let $x,z$ satisfy Assumption \ref{AssumptionCommon} and be zero mean.  Suppose there exists some $q\in [m,k]$  such that the matrix ${C^{x^Q \to z^Q }}(m,q)$,
	involving covariances of the data obtained by non-uniform quantizers, is full-rank. Then $x$ Granger causes $z$, provided that the  following condition is satisfied:
	\begin{align} \label{BoundNonUniformUsingInftyNorm}
	\sqrt{N_o N_I 	 } < \sigma_{\text{min}}^Q, 
	\end{align}
	where $N_o$ and $N_I$ are defined in \eqref{BoundForOneNorm_NonUniform} and \eqref{BoundForInfty_NonUniform},
	   $\sigma_{\text{min}}^Q$ is the smallest singular value of ${C^{x^Q \to z^Q }}(m,q)$ and $S_{xz}$,
	   $S_{zz}$ and $S_z$ are defined in \eqref{Easy_Sxz}-\eqref{Easy_Sz}.  \hfill $\square$
\end{prop}
\begin{rem}
Note that we do not make assumption on the identifiability of the second-order statistics of the unquantized jointly Gaussian signals through quantized measurements. The {\em a priori} information required to exploit Proposition \ref{NonUniformTheorem} are the upper bounds on the cross-correlation coefficients between signal $x$ and $z$ ($\overline \rho_{xz} $), the auto-correlation coefficients of  $z$  at nonzero lags ($\overline \rho_{zz} $), and the range of the standard deviation of $z$ ($\underline \gamma_z$ and $\overline \gamma_z $).
\end{rem}
\begin{rem}
Sufficient condition \eqref{BoundNonUniformUsingInftyNorm} is useful for investigating  Granger causality between jointly Gaussian signals through quantized observations. The RHS can be estimated from the quantized data available and the LHS of  the condition depends on the quantizer specifications and on prior  knowledge of bounds on the underlying jointly Gaussian statistics ($\overline \rho_{xz} $, $\overline \rho_{zz} $, $\underline \gamma_z $ and $\overline \gamma_z $). 
\end{rem}
\begin{rem} \label{SufficientCOnditionForAchiviebility_NonUniform}
It can be shown that \eqref{BoundNonUniformUsingInftyNorm} is satisfied if:
	\begin{align} \label{SuffCondBoundNonUniformUsingInftyNorm}
	\sqrt{N_o N_I	 } < \frac{\sigma_{\text{min}}}{2 }, 
	\end{align}
	where $\sigma_{\text{min}}$ is the smallest singular value of the causality matrix $C_G^{x \to z}(m,q)$. Note that the RHS of \eqref{SuffCondBoundNonUniformUsingInftyNorm}  does not depend on the quantization scheme and can be lower-bounded in terms of the prior bounds on the underlying statistics, if the underlying system is indeed causal. 
 The difference between the LHS and RHS can be thought of as a causality margin, which we would like to be as large as possible. This can be achieved by designing quantization parameters so that the LHS is as small as possible, subject to constraints on the number of levels available. This  leads to designs that may be very different from minimum mean square error (MMSE) quantizers. 
\end{rem}
 \begin{rem}
		The approaches described in Sections \ref{BinarySection} and \ref{GCQuantized} can be exploited for Granger causality inference of jointly Gaussian processes passing through other nonlinearities as well. Two cases can be considered. The first one is about nonlinearities whose relations between covariances or correlation coefficients of jointly Gaussian signals and second-order moments of output signals from the nonlinearity are invertible. For such situations, necessary and sufficient conditions for inferring Granger causality similar to  Section \ref{BinarySection} using Price's Theorem and Theorem \ref{MainTheoremOfCG} can be introduced. The second cases are nonlinearities where the identifiability is not guaranteed. Sufficient conditions to infer Granger causality between the jointly Gaussian signals using approaches similar to Section \ref{GCQuantized} can be derived. In fact, first the relationship between covariances of jointly Gaussian signals and output signals from nonlinearities through Price's Theorem can be obtained. Then bounds on the difference between covariances of jointly Gaussian signals and of the output signals from nonlinearities can be developed. And finally Theorem \ref{GeneralTheoremForQuantization} to derive sufficient conditions can be used.
\end{rem} 
\begin{rem}
	If the quantizers both have a threshold $c_i$ and $d_j$ at the origin, then an alternative approach is to lump the positive and negative quantization intervals together and then apply the binary quantizer results of section \ref{BinarySection}. However, the multilevel quantization approach of this section has the advantage that the empirical 
		 estimation of the covariances of the quantized signals is expected to become more accurate as the number of levels increases, see e.g. \cite{Thompson2017} and references therein for discussion.\\
			We further remark that 
			 by numerically inverting \eqref{CovNonUniformMultilevel1} and similar equations, the unquantized  variances and correlation coefficients can be recovered exactly from the  (co)variances of the quantized signals. In principle, causality or non-causality could then be determined from the rank of \eqref{DefinitionOfR}. However, when used on empirical estimates of the quantized signal statistics, this technique can be unreliable  since it can be shown that $\frac{\partial \rho}{\partial \gamma_{w_1^Q w_2^Q }}$ becomes arbitrary large if  the underlying  correlation coefficient $\rho$ is sufficiently strong. This 
			  is called the law of propagation of variance (or of statistical error); see e.g. \cite{Deming} for discussion.  In contrast, the approach taken in Proposition \ref{NonUniformTheorem} does not require inverting the functions and does not suffer from such issues.
\end{rem}
 \section{Granger causality using infinite-level uniformly  quantized data} \label{GCUnderInfiniteLevel}
In this Section,  Granger causality between jointly Gaussian signals is investigated when there is an infinite number of quantization levels.  In the following Subsection, we focus mainly on the high resolution regime where the resolution of the quantizers is sufficiently fine, i.e. $\Delta_z,\Delta_x\to 0$.  The case of infinite-level quantization with finite resolution is explored in Appendix \ref{GCUnderMidTread}. 

Infinite-level quantization has been studied in control systems \cite{EliaMitter2001}, system identification \cite{GustafssonKarlsson2009, RisuleoBottegal2020}, communications \cite{WidrowKollar}, etc. Although practical quantizers have a finite number of levels,  in the case of input signals with unbounded support they are difficult to analyse, because of the unbounded overload regions. In contrast, infinite-level quantizers have no overload regions to complicate the analysis. If the probability mass in the overload regions is sufficiently small, then an infinite-level quantizer is a reasonable approximation.

\subsection{High-resolution quantization} \label{GCUnderHRQ}

Recall that  $w_1, w_2$ denote $x$ and/or $z$ at different time lags and $\epsilon_1, \epsilon_2$  are the corresponding quantization error terms. By analysing the infinite sums representing covariances between quantization errors and signals for sufficiently small $k_i :=\frac{\Delta_{w_i}}{ \gamma_{w_i} }, i=1,2$,  it is shown in Appendices \ref{Big_Oh_Derivations_EW} and \ref{Big_Oh_Derivation_EE} that  the last three terms on the RHS of \eqref{RelationBetQandUnq} can be expressed as follows:
\begin{align}
	& |\gamma_{w_1 \epsilon_2}| < 4  |\gamma_{w_1 w_2} |   e^{-\frac{2\pi^2 }{k_2^2}}, \label{MainCovXEZeroMean7}\\
	&|\gamma_{ \epsilon_1 w_2} | < 4  |\gamma_{w_1 w_2} |   e^{-\frac{2\pi^2 }{k_1^2}} \label{MainCovEXZeroMean7},\\
	\gamma_{\epsilon_1 \epsilon_2} = &O\big( k_1 k_2 e^{-2\pi^2  (1- |\rho_{w_1 w_2}|) (\frac{1}{k_1^2} + \frac{1}{k_2^2})} \big), \label{MainCovEEZeroMean7}
\end{align}
where  $O(.)$ is Bachmann-Landau $O$-notation.  

Furthermore, the variance of the quantized signal  can be represented as follows  as shown in Appendix \ref{Big_Oh_Derivations_Var}:
\begin{align} \label{MainVar4}   
	\gamma_{w_1^Q}^2 = \gamma_{w_1}^2 + \frac{k_1^2 \gamma_{w_1}^2 (=\Delta_{w_1}^2)}{12} + O \Big(   e^{-\frac{ 2\pi^2}{k_1^2}} \Big).
\end{align}
In Appendix \ref{UpperBound_NormTwo_HR}, we  show that in the high resolution regime:
 \begin{align} \label{Norm_Two_HR_Final}
\| {\Gamma_{\epsilon}} (m,q) \|_2 <& 4 (m+1) \overline \gamma_{xz} e^{-\frac{2\pi^2}{k_x^2}} + \frac{\Delta_{z}^2}{12} + \nonumber\\
& \Big(4(m+1) \overline   \gamma_{xz}  e^{-\frac{2\pi^2}{k_x^2}}
+ \frac{\Delta_{z}^2}{12} \Big)^{\frac{1}{2}} O\bigg( \max\Big\{ e^{-\frac{ \pi^2}{k_z^2}},\nonumber\\
&  ( k_x k_z)^{\frac{1}{2}} e^{- \pi^2  (1- \overline \rho_{xz}) (\frac{1}{k_x^2} + \frac{1}{k_z^2})} 	 ,   k_z e^{-\frac{2\pi^2}{k_z^2}  (1-  \overline \rho_{zz} )  } \Big\} \bigg),
\end{align} 
where $\overline \gamma_{xz}$ is the upper bound on the cross-covariance between signals $x$ and $z$. The following result then follows immediately from Theorem \ref{GeneralTheoremForQuantization}:

\begin{prop} 
	\label{HighResProposition}
 	Let $x,z$ satisfy Assumption \ref{AssumptionCommon} and be zero mean.  Suppose there exists some $q\in [m,k]$ such that the matrix ${C^{x^Q \to z^Q }}(m,q)$
	\eqref{CQDefinition},  involving covariances of the high-resolution, uniformly quantized data, is full-rank. Then $x$ Granger causes $z$, provided that: 
	\begin{align}	\label{HighResProposition_Ineq}
4 (m+1) \overline \gamma_{xz} \exp \Big( -\frac{2\pi^2 \underline{\gamma}_x^2}{\Delta_x^2} \Big)  + \frac{\Delta_{z}^2}{12} +d(  \Delta_x,\Delta_z)  < \sigma_{\text{min}}^Q, 
	\end{align}
	where $\sigma_{\text{min}}^Q$ is the smallest singular value of ${C^{x^Q \to z^Q }}(m,q)$ and 
	\begin{align}
	d(  \Delta_x,\Delta_z  )&    :=\Big(4(m+1) \overline   \gamma_{xz}  e^{-\frac{2\pi^2}{k_x^2}}
	+ \frac{\Delta_{z}^2}{12} \Big)^{\frac{1}{2}} O\bigg( \max\Big\{e^{-\frac{ \pi^2}{k_z^2}}, \nonumber\\
	& \hspace{5mm}    ( k_x k_z)^{\frac{1}{2}} e^{- \pi^2  (1- \overline \rho_{xz}) (\frac{1}{k_x^2} + \frac{1}{k_z^2})} 	 ,   k_z e^{-\frac{2\pi^2}{k_z^2}  (1-  \overline \rho_{zz} )  } \Big\} \bigg).
	\end{align}  \hfill $\square$
\end{prop}

\begin{rem} \label{HighResRemark}
	This result gives an explicit formula in the high-resolution regime for deciding how finely quantized $x$ and $z$ should  be in order to infer causality from ${C^{x^Q \to z^Q }}(m,q)$, which can be constructed through quantized signals available. 
	As in  the quantization perturbation bound in Theorem \ref{GeneralTheoremForQuantization}, the RHS depends on the quantization scheme.
	Using \eqref{Achievability3} and \eqref{Norm_Two_HR_Final},  we can establish a quantizer-independent one hand side as follows:
	\begin{align} \label{FinalAchievability}
  8 (m+1) \overline \gamma_{xz} \exp \Big( -\frac{2\pi^2 \underline{\gamma}_x^2}{\Delta_x^2} \Big)  + \frac{\Delta_{z}^2}{6}  +2d(  \Delta_x,\Delta_z ) < \sigma_{\text{min}},
	\end{align}
	where $\sigma_{\text{min}}:=\sigma_{\text{min}}({C_G^{x \to z}}(m,q))$.\\
 Note the LHS  suggests that $\Delta_z$ plays a more critical role than $\Delta_x$. This is related to the fact that in the causality matrix $C_G^{x \to z}(m,q)$, $x$ appears only in cross-covariances with $z$, whereas $z$ also appears in auto-covariances with lag zero. Even if $\Delta_x$ is coarse, the inequality  can be satisfied by choosing a sufficiently  fine $\Delta_z$. Conversely,  if $\Delta_x$ is small,  $\Delta_z$  would still have to be small to satisfy \eqref{FinalAchievability}. Indeed it can be shown that for the LHS to be as small as possible, we require 
 	$1/\Delta_z \approx \exp \Big(  \frac{ \pi^2 \underline{\gamma}_x^2}{\Delta_x^2} \Big)$, i.e. the resolution in $z$ should be exponentially finer than the resolution in $x$. Furthermore, to improve the causality margin, the LHS can be minimized with respect to $\Delta_{x}$ and $\Delta_{z}$ under appropriate  constraints  on quantization resolution, e.g. 
 		$\log\frac{1}{\Delta_{x}}+ \log\frac{1}{\Delta_{z}} \leq R$   (which can be interpreted as the total expected bit rate in a variable-rate code for  quantized $x$ and $z$).  
\end{rem}

\section{Estimation of second-order statistics of quantized data} \label{EmpiricalEstimation-sec}

 In this section,  we establish that the auto- and cross-covariances between the quantized data can be consistently estimated, enabling the matrix ${C^{x^Q \to z^Q }}(m,q)$ to be constructed. To do so,  the ergodic theorem  is exploited. 
 
First, we state the following theorem. 
 \begin{thm}   \label{ErgodicityPreservation} 
		Let $(y_k)_{k \geq 1}$   be a stationary, ergodic $l$-variate vector process and  ${f}$  be  a 
	measurable  function  $ {f}  :  \mathbb{R}^{l \times \infty} \to \mathbb{R}^s$.  Let  $\xi_k  = {f}({y}_k, {y}_{k+1},...)$  define 
	an $s$-variate vector process $(\xi_k)_{k \geq 1}$.  Then  $(\xi_k)_{k \geq 1}$ is  stationary  ergodic. 
\end{thm}
\begin{pf}
 	See Appendix \ref{Proof_Of_ErgodicityPreservation}. It follows the same line of \cite{Breiman} with modifications to allow  vector processes. 
\end{pf} 
Theorem \ref{ErgodicityPreservation} implies that jointly Gaussian stationary, ergodic signals $x_k$ and $z_k$ remain stationary and ergodic after quantization ($x^Q_k$ and $z^Q_k$). Furthermore, the Theorem implies that   the products  $x^Q_kx^Q_{ k+\kappa }$, $z^Q_kz^Q_{ k+\kappa }$, $x^Q_kz^Q_{ k+\kappa }$ are also stationary and ergodic, for instance, by defining $y_k =[x_k, z_k]^\intercal$, $\xi_k = f(y_k,y_{k+1},...) = x^Q_k z^Q_{k+\kappa}$.  This enables us to exploit the Ergodic Theorem to estimate the covariance functions  of the quantized signals. 

\begin{thm} [Ergodic Theorem \cite{Breiman}] \label{ErgodicityTheorem}  Suppose $(\xi_k)_{k \geq 1}$ is a strict-sense stationary and ergodic univariate process with $E|\xi_1|<\infty$, then almost surely and in the first mean
	\begin{align} 
		\lim\limits_{n \to \infty} \frac{1}{n} \sum_{k=1}^{n}\xi_k = E \{\xi_1\}.
	\end{align} \hfill $\blacksquare$
\end{thm}
 In the following, $w_{1,k}^Q$ and $w_{2,k}^Q$ denote $x^Q_k$ and/or $z^Q_k$. 
 \begin{rem}
	The standing assumption in this Section is that the unquantized processes are ergodic.  For multivariate continuous-time Gaussian processes \cite{Adler} and univariate discrete-time Gaussian processes 
\cite{CramerLeadbetter}, it is known that ergodicity follows under stationarity if the auto- and cross-covariances vanish as the lag  approaches infinity. In Appendix \ref{Proof_of_Ergodocity_For_GP}, this result is extended to stationary multivariate discrete-time  Gaussian processes. 
\end{rem} 

\subsection{Auto- and cross-covariance estimators of quantized signals with zero mean}

In order to estimate the auto- and cross-covariance between quantized signals with mean zero, Theorem \ref{ErgodicityTheorem} can be used. For instance, the cross-covariance between quantized signals with mean zero $w_{1,k}^Q$ and $w_{2,k}^Q$ at lag $\kappa$ can be estimated as follows. Define $\xi_k := w_{1,k}^Q w_{2,k+\kappa}^Q$ and note that as discussed after Theorem \ref{ErgodicityPreservation}, $\xi_k$ is stationary and ergodic. Furthermore, it is clear that $E|\xi_1|$ is finite. The following almost surely  and  in the first mean convergent estimator can be presented:
\begin{align} 
	\frac{1}{n} \sum_{k=1}^{n-\kappa} w_{1,k}^Q w_{2, k+\kappa}^Q \to \gamma_{w_1^Qw_2^Q}(\kappa) := E \{w_{1,k}^Q w_{2,k+\kappa}^Q \}.  
\end{align}

\subsection{Auto- and cross-covariance estimators of quantized signals with  non-zero mean}

For  non-zero mean quantized signals, we can still use the theorem as follows. We know that:
 \begin{align} \label{NonZero1}
	\gamma_{w_1^Qw_2^Q}(\kappa) = E \Big\{w_{1,k}^Q w_{2, k+\kappa}^Q \Big\} - E\Big\{w_{1,k}^Q\Big\}E\Big\{w_{2,k}^Q \Big\}
\end{align} 
Each terms on the right hand side of \eqref{NonZero1} can be estimated through Theorem \ref{ErgodicityTheorem}. Let us first mention the following theorem:
\begin{thm} \cite{Davidson} \label{Borel1} Let  $h: \mathbb{R}^r \to \mathbb{R}$ be a Borel function, continuous at $a$. If $\eta_n \to a$ almost surely, then $h(\eta_n) \to h(a)$ almost surely.   \hfill $\blacksquare$
\end{thm}
Using Theorem \ref{Borel1}, with $r=3$ and $h(\eta_{1,n}, \eta_{2,n}, \eta_{3,n}) = \eta_{1,n} - \eta_{2,n} \eta_{3,n}$,  
 the following almost surely convergent estimator can be obtained: 
\begin{align} \label{Final1}
	\Bigg(\frac{1}{n} \sum_{k=1}^{n-\kappa} w_{1,k}^Q w_{2, k+\kappa}^Q - \frac{1}{n^2} \sum_{k=1}^{n} w_{1,k}^Q  \sum_{k=1}^{n}  w_{2, k}^Q \Bigg) \to \gamma_{w_1^Qw_2^Q}(\kappa).
\end{align}

\section{Simulation} \label{Simulation-sec}
 In this Section, numerical examples  are presented to illustrate the proposed methods. We consider the following second order processes from \cite{Gourevitch}:  
\begin{align}
	x_k &= 0.95 \sqrt{2} x_{k-1} - 0.9025 x_{k-2} - 0.9 z_{k-1} + 0.5 e_{1,k}+ 0.5 e_{{k}},\label{Simulation_x_k}\\
	z_k&= -1.05 z_{k-1} -0.85 z_{k-2} - 0.8 x_{k-1}+ 0.5 e_{2,k} + 0.5 e_{{k}}, \label{Simulation_z_k}
\end{align}
where $e_{1,k},e_{2,k}, e_{{k}}$ are assumed to be white Gaussian signals with zero mean and unit variance. 

\begin{figure}[!h]
 	\centering
 	\includegraphics[width=3.5in]{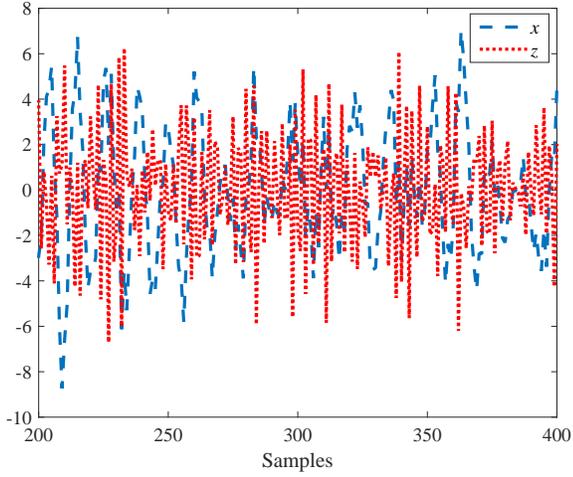}
 	\caption{Unquantized signals $z$ and $x$. }
 	\label{fig:Signalx}
 \end{figure}

We collect $n=1000$ samples from quantized versions of $x$  and $z$  to determine if  $x$ Granger causes $z$.  The unquantized signals $z$  and $x$  are depicted in Fig. \ref{fig:Signalx}   	to show how these signals change  with each other over the period of samples $[200,400]$. Note that we do not have access to such signals to investigate Granger causality. The process $z$ is partially Markov of order two. For noncausality cases, we change the coefficient of $ x_{k-1}$ in \eqref{Simulation_z_k}, i.e. $-0.8$, to zero.

 Note that empirical estimate of covariances at lag $\kappa$ becomes less reliable when the lag approaches  to the number $n$ of samples available. As a rule of thumb, it is suggested to estimate the covariances associated with at most a quarter of the number of samples ($n/4$) \cite{WeiTSA2005}.
\begin{figure}[!h]
	\centering
	\includegraphics[width=3.5in]{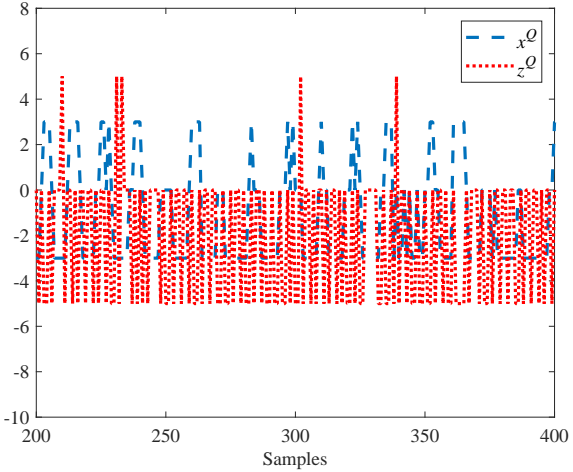}
	\caption{Quantized signals $z^Q$ and $x^Q$ with two-bit quantization.}
	\label{fig:Signalz}
\end{figure}
\subsection{Binary quantization}
 In this Subsection, the signals $x_k$ and $z_k$  are quantized by binary quantizers with the thresholds zero and the quantization levels $\{-1,1\}$. Using the estimators of the previous Section, we estimate the smallest singular value in Theorem \ref{CausalityForBinary} to be 0.4697. The value is significantly different from zero. Thus, it shows that the causality matrix is full rank and therefore, $x$ Granger causes $z$.  The true smallest singular value equals 0.4807, which is close to the above-mentioned estimated value. 
 
Now, we consider noncausality situation where the coefficient of $ x_{k-1}$ in \eqref{Simulation_z_k} is set to zero. In this situation, $x$ does not Granger cause $z$. To investigate noncausality through binary quantized observations, data are collected after binary quantization and rank of the causality matrix is determined to apply Theorem \ref{CausalityForBinary}.  The smallest singular value is 0.0095 which is near zero.   Hence, there is not strong evidence for the causality matrix to be full rank.  
  We would therefore conclude that $x$ does not Granger cause $z$.

 \subsection{Non-uniform finite-level quantization} 
We consider saturated quantizers with granular regions $[-3,3]$ and $[-5,5]$ associated, respectively,
 with $x$ and $z$, and equal quantization intervals in the granular regions. The  quantizer output is given by  the lower boundary point of the corresponding cell; i.e. in the cell $(c_i, c_{i+1}] $ the value of quantizer is $c_i$.   If the quantizer input falls outside the granular region, the quantizer takes the value of the nearest quantizer point as described above.  Let us consider that the variances are within intervals with widths 0.2 and upper bounds on auto-correlation coefficient  of signal $z$ and cross-correlation between signal $x$ and $z$ are, 0.7 and 0.5, respectively.

The corresponding quantized versions of signals $z$ and $x$  with two-bit quantization depicted in Fig. \ref{fig:Signalx} are shown in Fig. \ref{fig:Signalz}. The matrix constructed by the sample  covariances of the quantized signals satisfies the sufficient condition \eqref{BoundNonUniformUsingInftyNorm}, $\sqrt{N_o N_I 	 } - \sigma_{\text{min}}^Q  = -0.1346$ for $q=6$. From Proposition \ref{NonUniformTheorem}, we conclude that $x$ Granger causes $z$.  If the true covariances of the quantized signals through \eqref{CovNonUniformMultilevel1} are used to determine the causality, the difference between the right and left hand sides of the sufficient condition of Proposition \ref{NonUniformTheorem} is $-0.1480$ which is less than zero and it can be concluded that $x$ Granger causes $z$. 

Now let us set the coefficient of $ x_{k-1}$ in \eqref{Simulation_z_k}  zero and calculate the sufficient condition \eqref{BoundNonUniformUsingInftyNorm}. When the estimated and true covariances are used to construct the matrix $C^{x^Q \to z^Q}$, the values of $\sqrt{N_o N_I 	 } - \sigma_{\text{min}}^Q $ are, respectively, greater than   $+5.5820$ and $+5.5400$ for different values of $q$. The values are significantly bigger than zero and we cannot conclude that $x$ Granger causes $z$.

\begin{figure}[!h]
	\centering
	\includegraphics[width=3.5in]{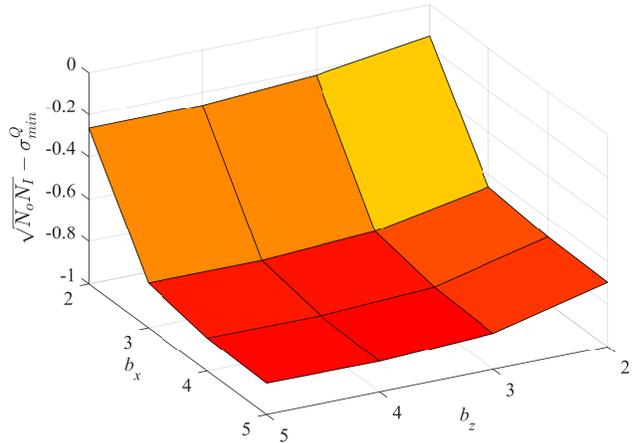}
	\caption{Values of sufficient condition \eqref{BoundNonUniformUsingInftyNorm} for different numbers of quantization bits. }
	\label{fig:LHS_sigmaminQ}
\end{figure} 
%
%
%
%
%
 Now we investigate how the numbers of bits denoted by $b_x$ and $b_z$ associated with quantizers of signals $x$ and $z$ impact the determination of  Granger causality using the approach given in Subsection \ref{NonuniformSection}. In  Fig. \ref{fig:LHS_sigmaminQ}, the sufficient condition \eqref{BoundNonUniformUsingInftyNorm}, i.e. $\sqrt{N_o N_I 	 } - \sigma_{\text{min}}^Q$  is depicted. As the numbers of bits increase, the perturbation reduces and the differences between LHS and RHS of \eqref{BoundNonUniformUsingInftyNorm}  becomes more significant.



\section{Conclusions}
\label{conclusions-sec}

In this paper, first we showed a rank-based representation for the equality of two conditional jointly Gaussian random vectors. Assuming joint Gaussianity and  a known  partial Markov order, we introduce the notion of a causality matrix and show that Granger causality is equivalent to this matrix having full rank. We also introduced a geometric interpretation for the smallest singular value of the causality matrix. The smallest singular value is indeed a measure of the distance between two conditional Gaussian distribution functions appearing in the probabilistic definition of Granger causality. We exploited such an interpretation to introduce  a new measure of causality.

 Next a necessary and sufficient condition was proposed for assessing Granger causality under binary quantization. Furthermore, conditions under which causality can be inferred by using just the statistical properties of the quantized signals instead of estimating the statistics or model of the underlying Gaussian signals were introduced. Such conditions have been proposed for a range of quantizers including non-uniform,  uniform and high resolution quantizers. Note that such an approach can be extended to other nonlinearities as well.
 
Future work will focus on  obtaining statistical confidence intervals and convergence rates for the empirical estimators proposed here.  Another open question involves relaxing the Gaussian assumption. In this case, it is likely that  information-theoretic approaches based on {\em directed information} (see e.g. \cite{Amblard}) will prove useful. In addition,  the verification of  Gaussianity   and  determination of the unknown partial Markov order through quantized
data are important topics for future study.    The estimation of rank and singular values has been extensively studied in the literature,  see e.g. \cite{Portier2014, Kleibergen2006}. For practical use, the statistical properties of such estimators for the matrices introduced in this paper will need  to be studied. We leave this as future work.   A rigorous study of statistical efficiency with quantized data is beyond the scope of the current paper. However, one possible approach is to exploit results from \cite{DandawateGiannakis1995} and verify its assumptions for quantized signals.





\appendix

\section{Proof of Theorem \ref{MainTheoremOfCGPDF}} \label{Proof_Of_MainTheoremOfCGPDF}
We begin by showing that the first two items in the Theorem are equivalent.
 
The conditional Gaussian random vectors $X|Z,Y$ and $X|Z,W$ have distributions as follows:
\begin{align*}
X|Z,Y &\sim \mathcal{N} ({\mu_\text{cond}^{X|Z,Y}}, {\Gamma_\text{cond}^{ X | Z,Y }}),\\
 X | Z,W  &\sim \mathcal{N} ({\mu_\text{cond}^{ X | Z,W }}, {\Gamma_\text{cond}^{ X | Z,W }}),
\end{align*}
where ${\mu_\text{cond}^{ X |Z,Y}}$ and ${\mu_\text{cond}^{ X | Z,W }}$ are conditional means and ${\Gamma_\text{cond}^{ X | Z,Y }}$ and ${\Gamma_\text{cond}^{ X | Z,W }}$ are conditional covariances of the random vectors $ X | Z,Y $ and $ X | Z,W $, respectively. It can be shown, e.g. \cite{KailathSayedHassibi2000}, that
\begin{align}
{\mu_\text{cond}^{ X | Z,Y }}:=& \mu_{ X } + \Gamma_{ X,Z } \Gamma_{ Z,Z }^{-1} ( Z -\mu_{ Z }) + (\Gamma_{ X,Y } - \Gamma_{ X,Z } \Gamma_{ Z,Z }^{-1}  \Gamma_{ Z,Y }) K_1 \times \nonumber\\
&\Big ( Y -\mu_{ Y } - \Gamma_{ Y,Z } \Gamma_{ Z,Z }^{-1} ( Z -\mu_{ Z })  \Big) \label{CondMeanAgivenCB}, \\
{\mu_\text{cond}^{ X | Z,W }}:=& \mu_{ X } + \Gamma_{ X,Z } \Gamma_{ Z,Z }^{-1} ( Z -\mu_{ Z }) + (\Gamma_{ X,W } - \Gamma_{ X,Z } \Gamma_{  Z,Z }^{-1}  \Gamma_{ Z,W }) K_2 \times \nonumber\\
&\Big ( W -\mu_{ W } - \Gamma_{ W,Z } \Gamma_{ Z,Z }^{-1} ( Z -\mu_{ Z })  \Big) \label{CondMeanAgivenCD},
\end{align}
where
\begin{align}
K_1:=& \big( \Gamma_{ Y,Y}  - \Gamma_{ Y,Z}   \Gamma_{ Z,Z }^{-1}      \Gamma_{ Z,Y }  \big)^{-1}, \label{DefK_1} \\
K_2:=& \big( \Gamma_{ W,W } - \Gamma_{ W,Z }  \Gamma_{ Z,Z }^{-1}      \Gamma_{ Z,W }  \big)^{-1}   \label{DefK_2}.
\end{align}
In order for two Gaussian distributions to be identical, it is necessary and sufficient that they should have the same mean vector and covariance matrix.  Let us begin with the equality of mean vectors (${\mu_\text{cond}^{ X | Z,Y }}={\mu_\text{cond}^{ X | Z,W }}$). Equating \eqref{CondMeanAgivenCB} and \eqref{CondMeanAgivenCD} and noting the positive definiteness of $\Gamma_{[ Y Z W ],[ Y Z W ]}$ imply that the coefficients of linear relationship between $ Y -\mu_{ Y }$, $  Z -\mu_{ Z }$ and $ W -\mu_{ W }$ appearing in ${\mu_\text{cond}^{ X | Z,Y }}={\mu_\text{cond}^{ X | Z,W }}$ must be zero. The coefficients are zero if and only if 
\begin{align}
& \Gamma_{ X,Y } - \Gamma_{ X,Z } \Gamma_{ Z,Z }^{-1}  \Gamma_{ Z,Y } =0 \label{RANK1}, \\
& \Gamma_{ X,W } - \Gamma_{ X,Z } \Gamma_{ Z,Z }^{-1}  \Gamma_{ Z,W } = 0 \label{RANK2}.
\end{align}
Thus, ${\mu_\text{cond}^{ X | Z,Y }}={\mu_\text{cond}^{ X | Z,W }}$ if and only if \eqref{RANK1} and \eqref{RANK2} hold. It can be shown that if \eqref{RANK1} and \eqref{RANK2} are satisfied, the equality of the conditional covariance matrices ${\Gamma_\text{cond}^{ X | Z,Y }}$ and ${\Gamma_\text{cond}^{ X | Z,W }}$ also holds. In other words, the conditional Gaussian probability density functions are equal ($P( X | Z,Y ) = P( X | Z,W )$) if and only if \eqref{RANK1} and \eqref{RANK2} are satisfied.

We know that a matrix $A$ is zero if and only if $\mathrm{rank} \hspace{.5mm}A =0$. Using Guttman rank additivity formula \cite{Guttman} for \eqref{RANK1} and \eqref{RANK2}, we have:
\begin{align} 
\mathrm{rank} \left[
\begin{array}{cc}
\Gamma_{ X,Y } & \Gamma_{ X,Z }\\
\Gamma_{ Z,Y } & \Gamma_{ Z,Z }
\end{array}
\right]  =\# Z  \label{RANK1_CONVERSE},\\
\mathrm{rank} \left[
\begin{array}{cc}
\Gamma_{ X,W } & \Gamma_{ X,Z }\\
\Gamma_{ Z,W } & \Gamma_{ Z,Z }
\end{array}
\right] =\# Z  \label{RANK2_CONVERSE}.
\end{align}
There are two partitioned matrices whose second column is the same in the equations above.  To deal with the ranks of these two matrices, let
\begin{align} \label{DefinitionOfXi}
	\Xi := 
	\left[
	\begin{array}{ccc}
		\Gamma_{ X,Y } & \Gamma_{ X,Z } & \Gamma_{ X,W }\\
		\Gamma_{ Z,Y } & \Gamma_{ Z,Z } & \Gamma_{ Z,W }
	\end{array}
	\right].
\end{align}
Using  Lemma 6 of \cite{LindquistBarrett}, it follows:
\begin{align} \label{UpperRank}
\mathrm{rank} \hspace{.5mm}
\Xi  \leq &
\mathrm{rank}
\left[
\begin{array}{cc}
\Gamma_{ X,Y } & \Gamma_{ X,Z  } \\
\Gamma_{ Z,Y } & \Gamma_{ Z,Z }
\end{array}
\right]  + \mathrm{rank}
\left[
\begin{array}{cc}
\Gamma_{ X,Z } & \Gamma_{ X,W }\\
\Gamma_{ Z,Z } & \Gamma_{  Z,W }
\end{array}
\right] \nonumber \\
& -\mathrm{rank}
\left[
\begin{array}{c}
\Gamma_{ X,Z } \\
\Gamma_{ Z,Z } 
\end{array}
\right] \nonumber \\
=& \# Z.
\end{align}
We also have:
\begin{align}\label{LowerRank}
\mathrm{rank} \hspace{.5mm}
\Xi  &\geq \max \Big \{\mathrm{rank}
\left[
\begin{array}{c}
\Gamma_{ X,Z } \\
\Gamma_{ Z,Z } 
\end{array}
\right], \mathrm{rank} \left[
\begin{array}{cc}
\Gamma_{ X,Y } &   \Gamma_{ X,W }\\
\Gamma_{ Z,Y } &  \Gamma_{ Z,W }
\end{array}
\right] \Big \}   \nonumber\\
& \geq  \# Z.
\end{align}
And it follows that:
\begin{align} \label{RankCond}
\mathrm{rank} \hspace{.5mm} \Xi  =\#Z.
\end{align}
We can conclude that if the conditional Gaussian probability density functions are equal ($P( X | Z,Y ) = P( X | Z,W )$), \eqref{RANK1} and \eqref{RANK2} are satisfied and in turn show that the rank criterion \eqref{RankCond} holds. Therefore, one direction is proved.

In order to prove the converse, we should show that \eqref{RANK1_CONVERSE} and \eqref{RANK2_CONVERSE} are satisfied if \eqref{RankCond} holds. Note that since $\Gamma_{ Z,Z }$ is positive definite, \eqref{RankCond} implies that both \eqref{RANK1_CONVERSE} and \eqref{RANK2_CONVERSE} hold.

\textit{Proof of item 3:} Here we show that items 1 and 3 are equivalent. If item 1 in the Theorem holds, we know that \eqref{RANK1_CONVERSE} and \eqref{RANK2_CONVERSE} (and therefore \eqref{RANK1} and \eqref{RANK2}) are satisfied. Thus, \eqref{CondMeanAgivenCB} and \eqref{CondMeanAgivenCD} are both equal to ${\mu_\text{cond}^{ X | Z,Y }} = {\mu_\text{cond}^{ X | Z,W }} = \mu_{ X } + \Gamma_{ X,Z } \Gamma_{ Z,Z }^{-1} ( Z  -\mu_{ Z })$, where $ \mu_{ X } + \Gamma_{ X,Z } \Gamma_{ Z,Z }^{-1} ( Z  -\mu_{ Z })$ is the conditional mean of $ X $ given $ Z $, $\mu_\text{cond}^{ X | Z }$. It 
follows that ${\Gamma_\text{cond}^{ X | Z,Y }} = {\Gamma_\text{cond}^{ X | Z,W }} = {\Gamma_\text{cond}^{ X | Z }}$, where $ {\Gamma_\text{cond}^{ X | Z }}$ is the conditional covariance of $ X $ given $ Z$. This completes the proof.

\section{Proof of Theorem \ref{MainTheoremOfCG}} \label{Proof_Of_MainTheoremOfGNC}
We can apply Theorem \ref{MainTheoremOfCGPDF} to   \eqref{PracticalGCDef} with $ X :=z_{k+1}$, $ Y := x^k_{k-m+1}$, $ Z :=z^k_{k-m+1}$ and $ W  :=z^{k-m}$,
to prove the first two parts of the Theorem.\\
To prove the last part, let us also define $ T := x^{k-m} $. We have:
\begin{align}
E( X | Z,W ) =& E\Big(E( X |  Y,Z,W,T ) \Big|  Z,W \Big) \nonumber\\
=& E\Big(E( X | Y,Z ) \Big| Z,W   \Big) \nonumber\\
=& E\Big(E( X | Z ) \Big|  Z,W  \Big)\nonumber\\
=& E( X | Z ),
\end{align}
where the first equality follows from the law of iterated expectations, the second equality follows from Assumption \ref{AssumptionOnOrderOFJoint} and the third one is due to the assumption of the last part of the Theorem. The conditional variance of $ X | Z,W $ equals the conditional variance of $ X | Z $ as well. This completes the proof.

\section{Proof of Theorem \ref{LowerBoundOnDifferenceBetweenConditionalMeans}} \label{ProofOfGeometric}
   The variance  of the difference between ${\mu_\text{cond}^{X|Z,Y}}$ and ${\mu_\text{cond}^{X|Z,W}}$ can be upper and lower bounded using Rayleigh quotient theorem \cite{horn2013matrix} as follows:
\begin{align} \label{Finale1}
	\lambda_{\text{min}} (J) \Bigg\| \left[\begin{array}{c}  
		a \\
		b
	\end{array}\right] \Bigg\|_2^2	\leq \text{Var} &\Big({\mu_\text{cond}^{X|Z,Y}} - {\mu_\text{cond}^{X|Z,W}}\Big)\leq \lambda_{\text{max}} (J) \times \nonumber\\
	&\hspace{20mm}  \Bigg\| \left[\begin{array}{c}  
	a \\
		b
	\end{array}\right] \Bigg\|_2^2,
\end{align}
where 
\begin{align}
	J:=& \left[\begin{array}{cc}
		K_1  &   K_3 \\
		K_3^\intercal & K_2
	\end{array}\right],
\end{align}
 $K_1$, $K_2$  are defined in \eqref{DefK_1}-\eqref{DefK_2} and
\begin{align} 
	K_3:=&-K_1 \big(  \Gamma_{Y,W} - \Gamma_{Y,Z} \Gamma_{Z,Z}^{-1}  \Gamma_{Z,W} \big) K_2, \\
 a := & (\Gamma_{X,Y} - \Gamma_{X,Z} \Gamma_{Z,Z}^{-1}  \Gamma_{Z,Y})^\intercal   \label{DefA}, \\
 b := & ( \Gamma_{X,W} - \Gamma_{X,Z} \Gamma_{Z,Z}^{-1}  \Gamma_{Z,W} )^\intercal  \label{DefB}.
\end{align}
Note that $X$ is a scalar random variable in this Theorem therefore $a$ and $b$ are column vectors. We have:
\begin{align}
	\text{Var} \Big({\mu_\text{cond}^{X|Z,Y}} - {\mu_\text{cond}^{X|Z,W}}\Big)  \geq& \lambda_{\text{min}} (J) \Bigg\| \left[\begin{array}{c}  
		a\\
		b
	\end{array}\right] \Bigg\|_2^2 \nonumber\\
	=& \lambda_{\text{min}} (J)  \|	c   \|_2^2 \nonumber\\
	=& \lambda_{\text{min}} (J)   c^\intercal c  \nonumber\\
	=& \lambda_{\text{min}} (J)  \sigma_{\max}^2(c),
\end{align}
where
\begin{align}
	c:=\Big(   \left[\begin{array}{cc}  
		\Gamma_{X,Y}  & 	\Gamma_{X,W}   
	\end{array}\right]  - \Gamma_{X,Z} \Gamma_{Z,Z}^{-1}   \left[\begin{array}{cc}  
		\Gamma_{Z,Y}  & 	\Gamma_{Z,W}   
	\end{array}\right]  	\Big)^\intercal.
\end{align}
Using Corollary 2.9 of \cite{ZhangSchur}, we have:
\begin{align}
	\sigma_{\max}^2	(c	) \geq& \sigma_{\min}^2 \Bigg( \left[\begin{array}{ccc}
		\Gamma_{Z,Z}  & \Gamma_{Z,Y}  & \Gamma_{Z,W} \\
		\Gamma_{X,Z}  & \Gamma_{X,Y}  & \Gamma_{X,W} 
	\end{array}\right]\Bigg)  \nonumber\\
	=& \sigma_{\min}^2 \Bigg(\left[\begin{array}{ccc}
		\Gamma_{X,Y}  & \Gamma_{X,Z}  & \Gamma_{X,W} \\
		\Gamma_{Z,Y}  & \Gamma_{Z,Z}  & \Gamma_{Z,W} 
	\end{array}\right]\Bigg).
\end{align}
Therefore, the lower bound on the variance between random variables  ${\mu_\text{cond}^{X|Z,Y}}$ and ${\mu_\text{cond}^{X|Z,W}}$  is as follows:
\begin{align} \label{Final2}
	\lambda_{\text{min}} (J)  \sigma_{\min}^2 \Bigg( \left[\begin{array}{ccc}
		\Gamma_{X,Y}  & \Gamma_{X,Z}  & \Gamma_{X,W} \\
		\Gamma_{Z,Y}  & \Gamma_{Z,Z}  & \Gamma_{Z,W} 
	\end{array}\right] \Bigg) \leq \text{Var} \Big({\mu_\text{cond}^{X|Z,Y}} - {\mu_\text{cond}^{X|Z,W}}\Big) .
\end{align}
Furthermore, the smallest eigenvalue of the matrix $J$ is always positive since the matrix $J$ can be written as:
\begin{align}
	J &= \left[\begin{array}{cc}
		K_1  &   0  \\
		0 & K_2
	\end{array}\right]  \left[\begin{array}{ccc}
		I &   - \Gamma_{Y,Z} \Gamma_{Z,Z}^{-1} &  0  \\
		0 &  \Gamma_{W,Z} \Gamma_{Z,Z}^{-1}   & -I
	\end{array}\right]
	\Gamma_{[Y Z W],[Y Z W]} \times \nonumber\\
	&\hspace{20mm}\left[\begin{array}{cc}
		I &   0  \\
		-\Gamma_{Z,Z}^{-1} \Gamma_{Y,Z}^\intercal  &   \Gamma_{Z,Z}^{-1}  \Gamma_{W,Z}^\intercal\\  0   & -I
	\end{array}\right]    \left[\begin{array}{cc}
		K_1  &   0  \\
		0 & K_2
	\end{array}\right] ,
\end{align}
and the covariance matrix of the random vector $[Y\ \ Z \ \ W]$ is positive definite by the assumption of the Theorem.

Now let us find a lower bound on $ \lambda_{\min} (J)$. First note that $J$ can be expressed as follows:
\begin{align}
	J = \left[\begin{array}{cc}
		K_1  &   0  \\
		0 & -K_2
	\end{array}\right]  J' \left[\begin{array}{cc}
		K_1  &   0  \\
		0 & -K_2
	\end{array}\right] ,
\end{align}
where
\begin{align}
 J':=	\left[\begin{array}{ccc}
		\Gamma_{Y,Y} - \Gamma_{Y,Z} \Gamma_{Z,Z}^{-1}  \Gamma_{Z,Y} &   \Gamma_{Y,W} -  \Gamma_{Y,Z}  \Gamma_{Z,Z}^{-1} \Gamma_{Z,W} \\
		\Gamma_{W,Y} -   \Gamma_{W,Z} \Gamma_{Z,Z}^{-1}   \Gamma_{Z,Y}  & \Gamma_{W,W} -  \Gamma_{W,Z}  \Gamma_{Z,Z}^{-1} \Gamma_{Z,W}
	\end{array}\right]  .
\end{align}
Note that the   matrix $J'$  is positive definite since  $\Gamma_{[YZW],[YZW]}$ is positive definite. Furthermore, for a square matrix $A$ and a positive definite matrix $B$ it can be shown $\lambda_{\min}(ABA) \geq \lambda_{\min}(B)   \lambda_{\min} (AA)$ using the definition of eigenvalue and Rayleigh quotient theorem. Therefore, we can have:
 \begin{align} \label{EigenValueJ_1}
	\lambda_{\min} (J) &\geq \lambda_{\min} \Big(   \left[\begin{array}{cc}
		K_1 ^2   &   0  \\
		0 & K_2 ^2 
	\end{array}\right]  \Big)  \lambda_{\min} (J') \nonumber\\
&=  \lambda_{\min}^2 \Big(   \left[\begin{array}{cc}
	K_1   &   0  \\
	0 & K_2 
\end{array}\right]   \Big)  \lambda_{\min} (J') .
\end{align} 
Moreover, we have:
\begin{align} 
	\lambda_{\min} (J') &=  \lambda_{\min} \Big( \Gamma_{[Y W], [Y W]}  -  
	\left[\begin{array}{c}
		\Gamma_{Y,Z}  \\
		\Gamma_{W,Z} 
	\end{array}\right]  \Gamma_{Z,Z}^{-1} \left[\begin{array}{cc}
		\Gamma_{Z,Y}  & \Gamma_{Z,W} 
	\end{array}\right]  \Big) \nonumber\\
&\geq    \lambda_{\min} \big( \Gamma_{[YZW], [YZW]} \big), \label{EigenValueJ_2}
\end{align}
 where the inequality above follows from  Corollary 2.9 of \cite{ZhangSchur}. 
 
Using the last two expressions above, we have:
\begin{align} \label{EigenValueJ_3}
	\lambda_{\min} (J) &\geq  \min\{ \lambda_{\min}^2 (K_1),  \lambda_{\min}^2 (K_2) \} \lambda_{\min} \big( \Gamma_{[YZW],[YZW]} \big) \nonumber\\
	&= \min\big \{ \lambda_{\max}^2 \big( K_1^{-1} \big),  \lambda_{\max}^2 \big( K_2^{-1}   \big)  \big \}    \lambda_{\min}\big( \Gamma_{[YZW],[YZW]} \big) \nonumber\\
	&\geq \min\bigg\{ \sigma_{\#Z+1}^2 \big(\Gamma_{[ZY],[ZY]}  \big), \sigma_{\#Z+1}^2 \big(\Gamma_{[ZW],[ZW]} \big)  \bigg\} \times \nonumber\\
	&\hspace{5mm}\lambda_{\min} \big( \Gamma_{[YZW],[YZW]} \big) \nonumber\\
	&= \min\big\{ \lambda_{\#Z+1}^2 \big( \Gamma_{[ZY],[ZY]}    \big),  \lambda_{\#Z+1}^2 \big( \Gamma_{[ZW],[ZW]}\big)  \big\} \times \nonumber\\
	&\hspace{5mm}\lambda_{\min} \big( \Gamma_{[YZW],[YZW]} \big),
\end{align}
where the second inequality above follows from the fact that for    positive definite symmetric matrices each eigenvalue equals its corresponding singular value  and from Corollary 2.9 of \cite{ZhangSchur}.
 And expression \eqref{FINAL_1_Theorem_DifCondMeans} follows from  \eqref{Final2} and \eqref{EigenValueJ_3}.

\section{Derivation of equation (\ref{CovNonUniformMultilevel1})}    \label{Derivation_Of_Eq_CovNonUniformMultilevel1}
Following \cite{Hagen1973, BenkevitcArXiv}, we derive here the relationship between covariances of jointly Gaussian random variables quantized by two different non-uniform quantizers and correlation coefficients of the underlying jointly Gaussian random variables.

Let us consider $g(w_1,w_2) = Q_1(w_1) Q_2(w_2)$ and $n=1$ in Price's Theorem.  We have:
\begin{align}
	\frac{\partial Q_1(w_1)}{\partial w_1} &= \sum_{i=1}^{n_1-1}(l_{i+1}-l_{i})\delta(w_1-c_{i}),\\
	\frac{\partial Q_2(w_2)}{\partial w_2} &= \sum_{j=1}^{n_2-1}(l'_{j+1}-l'_{j})\delta(w_2-d_{j}),
\end{align} 
where $\delta(.)$ is Dirac delta function. The RHS of \eqref{Price0} can be written as follows:
\begin{align} 
	&E\Big\{\frac{\partial Q_1(w_1)}{\partial w_1}.\frac{\partial Q_2 (w_2 )}{\partial w_2}\Big\} \nonumber \\
	&= \int\limits_{-\infty}^{+\infty} \int\limits_{-\infty}^{+\infty}\frac{\partial Q_1(w_1)}{\partial w_1} \frac{\partial Q_2(w_2)}{\partial w_2} f(w_1,w_2) dw_1 dw_2\nonumber \\
	&=\sum_{i=1}^{n_1-1}\sum_{j=1}^{n_2-1}(l_{i+1}-l_{i})(l'_{j+1}-l'_{j}) \times \nonumber \\
	&\qquad\qquad\quad \int\limits_{-\infty}^{+\infty} \int\limits_{-\infty}^{+\infty}\delta(w_1-c_{i})\delta(w_2-d_{j}) f(w_1,w_2) dw_1 dw_2 \nonumber \\
	&=\sum_{i=1}^{n_1-1}\sum_{j=1}^{n_2-1}(l_{i+1}-l_{i})(l'_{j+1}-l'_{j}) f(c_i,d_j).
\end{align}    
Price's Theorem implies that:
\begin{align*} 
	\frac{\partial E\Big\{ w_1^Q w_2^Q \Big \}}{\partial \gamma_{w_1 w_2}} = \sum_{i=1}^{n_1-1}\sum_{j=1}^{n_2-1}(l_{i+1}-l_{i})(l'_{j+1}-l'_{j}) f(c_i,d_j).
\end{align*} 
We have  from  the fundamental theorem of calculus:
\begin{align} 
	&E\Big\{ w_1^Q w_2^Q \Big \} - E\Big\{ w_1^Q w_2^Q \Big \} \Big |_{\gamma_{w_1w_2}^{ini}} \nonumber\\
	&= \sum_{i=1}^{n_1-1}\sum_{j=1}^{n_2-1}(l_{i+1}-l_{i})(l'_{j+1}-l'_{j})  \int\limits_{\gamma_{w_1w_2}^{ini}}^{\gamma_{w_1w_2}} f(c_i,d_j) d\gamma_{w_1w_2}.
\end{align}
Let us set $\gamma_{w_1w_2}^{ini}=0$ which implies that:
\begin{align} 
	E\Big\{ w_1^Q w_2^Q \Big \} \Big |_{\gamma_{w_1w_2}^{ini}=0} = E\{ w_1^Q \}E\{ w_2^Q \}.
\end{align}
Therefore, we have:
\begin{align} 
	& E\Big\{ w_1^Q w_2^Q\Big \}  - E\{ w_1^Q \}E\{ w_2^Q \} \nonumber \\
	&= \sum_{i=1}^{n_1-1}\sum_{j=1}^{n_2-1}(l_{i+1}-l_{i})(l'_{j+1}-l'_{j})  \int\limits_{0}^{\gamma_{w_1w_2}} f(c_i,d_j) d\gamma_{w_1w_2}.
\end{align}  
Thus, the covariance of quantized signals can be obtained as follows:
\begin{align} \label{Deriving_CovNonUniformMultilevel1}
	\gamma_{w_1^Q w_2^Q } :=& E\Big\{ w_1^Q w_2^Q \Big \} - E\{ w_1^Q \} E\{ w_2^Q \} \nonumber\\
	=&\sum_{i=1}^{n_1-1}\sum_{j=1}^{n_2-1}(l_{i+1}-l_{i})(l'_{j+1}-l'_{j}) \times \nonumber\\
	& \int\limits_{0}^{\gamma_{w_1w_2}} \frac{e^{\frac{-1}{2(1-\rho^2)}\big(\frac{c_i^2}{\gamma_{w_1}^2}-2 \rho \frac{c_i d_j}{\gamma_{w_1} \gamma_{w_2}}+\frac{d_j^2}{\gamma_{w_2}^2}\big)} }{2\pi\gamma_{w_1} \gamma_{w_2}\sqrt{1-\rho^2}}  d\gamma_{w_1w_2}    \nonumber \\
	=&\sum_{i=1}^{n_1-1}\sum_{j=1}^{n_2-1}(l_{i+1}-l_{i})(l'_{j+1}-l'_{j}) \times \nonumber\\
	& \int\limits_{0}^{\rho} \frac{1}{2\pi\sqrt{1-y^2}} e^{\frac{-1}{2(1-y^2)}\big(\frac{c_i^2}{\gamma_{w_1}^2}-2 y \frac{c_i d_j}{\gamma_{w_1} \gamma_{w_2}}+\frac{d_j^2}{\gamma_{w_2}^2}\big)}  dy,  
\end{align}                                          
which is \eqref{CovNonUniformMultilevel1}.

\section{Bounds on perturbation of covariances (Bounds on $S_{xz}$ and $S_{zz}$)} \label{PerturbationBound}
To facilitate dealing with the integral \eqref{CovNonUniformMultilevel1} appearing in  \eqref{Easy_Sxz} and \eqref{Easy_Szz},  an upper bound on the one-dimensional integrals can be obtained and then upper bounds on $S_{xz}$ and on $S_{zz}$ can be derived. In the following, we first find upper bounds on the function in integrand associated with \eqref{CovNonUniformMultilevel1} and then the integral and the maximum of the resulted upper bound are obtained to have upper bounds on $S_{xz}$ and on $S_{zz}$. \\
For ease of notation, let us first use $w_1$ and $w_2$ to denote $x$ and/or $z$ in the following. The final bounds are presented for signal $x$ and $z$ at the end of this Section for convenience as well.\\
The exponential function in \eqref{CovNonUniformMultilevel1} can be bounded as follows:
\begin{align}\label{ExpRelatedtoGaussian}
	&	e^{\frac{-1}{2(1-y^2)}\big(\frac{c_i^2}{\gamma_{w_1}^2}-2 y \frac{c_i d_j}{\gamma_{w_1} \gamma_{w_2}}+\frac{d_j^2}{\gamma_{w_2}^2}\big)}  \leq e^{-\frac{1}{2} \big(  \frac{c_i^2}{\gamma_{w_1}^2} + \frac{d_j^2}{\gamma_{w_2}^2} \big) +\frac{ \overline{\rho} }{1-{ \overline{\rho}}^2 } |\frac{c_i d_j}{\gamma_{w_1} \gamma_{w_2}}| }, \\
	&	e^{\frac{-1}{2(1-{ \overline{\rho}}^2)} \big(  \frac{c_i^2}{\gamma_{w_1}^2} + \frac{d_j^2}{\gamma_{w_2}^2}  + 2   \overline{\rho} |\frac{c_i d_j}{\gamma_{w_1} \gamma_{w_2}}| \big)  }    \leq e^{\frac{-1}{2(1-y^2)}\big(\frac{c_i^2}{\gamma_{w_1}^2}-2 y \frac{c_i d_j}{\gamma_{w_1} \gamma_{w_2}}+\frac{d_j^2}{\gamma_{w_2}^2}\big)}.
\end{align}
Let us define:
\begin{align} \label{DefKL}
	&K^L :=  \sum_{i=1}^{n_1-1}\sum_{j=1}^{n_2-1} \Delta_{l_i} \Delta_{l'_j} 
	e^{\frac{-1}{2(1-\overline{\rho}^2)} \big(  \frac{c_i^2}{\underline  \gamma_1^2} + \frac{d_j^2}{\underline  \gamma_2^2}  + 2 \overline{\rho}  \frac{|c_i d_j|}{ \underline  \gamma_1 \underline  \gamma_2  } \big) },\\
	&K^U := \max_{\substack{\underline \gamma_i \leq \gamma_{w_i}\leq \overline \gamma_i  \\ i=1,2}}  \sum_{i=1}^{n_1-1}\sum_{j=1}^{n_2-1} \Delta_{l_i} \Delta_{l'_j} 	e^{-\frac{1}{2} \big(  \frac{c_i^2}{\gamma_{w_1}^2} + \frac{d_j^2}{\gamma_{w_2}^2} \big) +\frac{ \overline{\rho}}{1-{ \overline{\rho}}^2 } \frac{| c_i d_j |}{\gamma_{w_1} \gamma_{w_2}} } \label{DefKU},
\end{align}
where $ \Delta_{l_i} := l_{i+1}-l_{i}$ and $\Delta_{l'_j} := l'_{j+1}-l'_{j} $.\\
Applying \eqref{ExpRelatedtoGaussian} - \eqref{DefKU} to \eqref{CovNonUniformMultilevel1}, the bounds on $\gamma_{w_1^Q w_2^Q }$ can be found as follows:
\begin{align} \label{Clarification1}
	\frac{1 }{2\pi}  K^L \arcsin(\rho) \leq \gamma_{w_1^Q w_2^Q } \leq 	\frac{1  }{2\pi} K^U\arcsin(\rho), \ \ \text{for}  \ \ \rho\geq 0, \\
	\frac{1 }{2\pi}  	K^U   \arcsin(\rho)   \leq \gamma_{w_1^Q w_2^Q }\leq \frac{1 }{2\pi}   K^L  \arcsin(\rho), \ \ \text{for}  \ \ \rho\leq 0.  
\end{align}
Thus the bound on $\gamma_{w_1^Q w_2^Q } - \gamma_{w_1w_2}$ is as follows:\\
For $\rho \geq 0$:
\begin{align} 
	\frac{K^L  }{2\pi} \arcsin(\rho) -\rho \gamma_{w_1} \gamma_{w_2} 
	\leq  &\gamma_{w_1^Q w_2^Q } -\gamma_{w_1w_2}  \nonumber\\ 
	& \hspace{3mm}\leq  \frac{K^U  }{2\pi}   \arcsin(\rho) -\rho \gamma_{w_1} \gamma_{w_2}, \label{CovNonUniformMultilevel_Bound_Rayleigh_FinalUpperBound1}
\end{align} 
for $\rho \leq 0$:
\begin{align}
	\frac{K^U  }{2\pi}   \arcsin(\rho) -\rho \gamma_{w_1} \gamma_{w_2} \leq & \gamma_{w_1^Q w_2^Q } - \gamma_{w_1w_2}   \nonumber\\ 
	& \hspace{3mm} \leq \frac{K^L  }{2\pi} \arcsin(\rho)  -\rho \gamma_{w_1} \gamma_{w_2}. \label{CovNonUniformMultilevel_Bound_Rayleigh_FinalUpperBound2} 
\end{align} 
In order to find the supremum of $|\gamma_{w_1^Q w_2^Q } - \gamma_{w_1w_2}|$, we have to find the maximum value of its upper bounds and minimum value of its lower bounds. Note that the upper bound and the lower bound are odd functions in $\rho$. Therefore, we just examine the maximum of $|K^U \arcsin(\rho) -\rho \gamma_{w_1} \gamma_{w_2}|$ and the minimum of $|	K^L  \arcsin(\rho) -\rho \gamma_{w_1} \gamma_{w_2}|$ for positive values of $\rho$ in the following. \\
The supremum of  $|\gamma_{w_1^Q w_2^Q } - \gamma_{w_1 w_2 }|$ is as follows:
\begin{align} \label{SupOfAbsoluteDifference_Rayleigh}
	\mathcal{S}_{w_1w_2} :=&\sup |\gamma_{w_1^Q w_2^Q } - \gamma_{w_1 w_2 }| \nonumber\\
	=&\max \big \{ U_{w_1w_2} , L_{w_1w_2} \big \},
\end{align}
where
\begin{align}
	U_{w_1w_2} &:=\max_{\substack{ 0 \leq\rho \leq  \overline{\rho} \\
			\underline \gamma_1 \underline \gamma_2   \leq \gamma'\leq \overline{\gamma}_1 \overline{\gamma}_2}} \Big| \frac{K^U  }{2\pi}    \arcsin(\rho) -\rho \gamma' \Big|, \\
	L_{w_1w_2} &:=\min_{\substack{ 0 \leq\rho \leq  \overline{\rho}\\
			\underline \gamma_1  \underline \gamma_2   \leq \gamma'\leq \overline{\gamma}_1 \overline{\gamma}_2 }} \Big |\frac{K^L  }{2\pi}   \arcsin(\rho) -\rho \gamma' \Big|	.
\end{align}
For the signals $x$ and $z$ and their quantized versions, the derived bounds above can be expressed as follows. Note that
\begin{align}
	|\gamma_{x^Qz^Q}(\kappa) - \gamma_{xz}(\kappa)| &\leq   \mathcal{S}_{xz},   \label{DifInCausalityMatrixElementsHRQ1} \\
	|\gamma_{z^Qz^Q}(\kappa) - \gamma_{zz}(\kappa)| &\leq  \mathcal{S}_{zz}, \ \ \kappa\neq 0, \label{DifInCausalityMatrixElementsHRQ2} 
\end{align}
where 
\begin{align}
	\mathcal{S}_{wz} &:= \max \{ U_{wz} , L_{wz}  \},    \label{SzzDef}
\end{align}
$w$ can be $x$ or $z$,
\begin{align}
	&	U_{wz} := \max_{\substack{0 \leq\rho \leq \overline{\rho}_{wz} \\
			\underline \gamma_w \underline \gamma_z   \leq \gamma'\leq \overline{\gamma}_w \overline{\gamma}_z}} \Big | \frac{K^U_{wz} }{2\pi} \arcsin(\rho) -\rho \gamma' \Big |,\\
	&	L_{wz} := \min_{\substack{0 \leq\rho \leq \overline{\rho}_{wz} \\
			\underline \gamma_w  \underline \gamma_z   \leq \gamma'\leq \overline{\gamma}_w \overline{\gamma}_z}} \Big| \frac{K^L_{wz}}{2\pi}  \arcsin(\rho) -\rho \gamma' \Big|,\\
	&	K^U_{wz} := \max_{\substack{\underline \gamma_w  \leq \gamma_{w}\leq \overline{\gamma}_w \\ \underline \gamma_z  \leq \gamma_{z}\leq \overline{\gamma}_z }}   \sum_{i=1}^{n_w-1}\sum_{j=1}^{n_z-1}{\Delta_w}_i   {\Delta_z}_j	e^{-\frac{1}{2} \big(  \frac{{c_i^w}^2}{\gamma_{w}^2} + \frac{{c_j^z}^2}{\gamma_{z}^2} \big) +\frac{\overline{\rho}_{wz} }{1-{\overline{\rho}_{wz} }^2 } \frac{| c_i^w c_j^z|}{\gamma_{w} \gamma_{z}}},\\
	&	K^L_{wz}:= \sum_{i=1}^{n_w-1} \sum_{j=1}^{n_z-1} {\Delta_w}_i   {\Delta_z}_j e^{\frac{-1}{2(1-{\overline{\rho}_{wz} }^2)} \big(  \frac{{c_i^w}^2}{\underline  \gamma_w^2} + \frac{{c_j^z}^2}{\underline  \gamma_z^2}  + 2 \overline{\rho}_{wz}   \frac{|{c_i^w} {c_j^z}| }{\underline  \gamma_w \underline  \gamma_z }  \big)},
\end{align}
$ \Delta_{w_i} := l_{i+1}^w-l_{i}^w$ and ${\Delta_z}_j := l_{j+1}^z-l_{j}^z $.

\section{Finite resolution quantization} \label{GCUnderMidTread}
We develop an upper bound on the norm of perturbation matrix caused by quantization and use Theorem \ref{GeneralTheoremForQuantization} to introduce a sufficient condition to infer Granger causality between jointly Gaussian signals $x$ and $z$ using uniform quantized signals.
To do so, we first derive an upper bound on the difference between covariances of quantized and unquantized signals in Appendix  \ref{Upper_Bounds_On_Difference_Covariances_Mid_Tread}.  Unlike the previous Section, we use results from \cite{WidrowKollar} on the cross-covariances of bivariate Gaussian random vectors that are subjected to infinite-level uniform quantization.  
\begin{prop} 
	\label{MidTreadProposition}
	Let $x,z$ satisfy Assumption \ref{AssumptionCommon} and be zero mean.  Suppose there exists some $q\in [m,k]$  such that the matrix ${C^{x^Q \to z^Q }}(m,q)$,
	involving covariances of the uniformly quantized data with $\Delta_x < 2\pi \underline \gamma_x$ and $\Delta_z < 2\pi \underline \gamma_z$, is full-rank.  Then $x$ Granger causes $z$, provided that the following condition is satisfied:
	\begin{align}
		\sqrt{N_{\infty} \hspace{1mm} N_1} &< \sigma_{\text{min}}^Q,\label{corollary_Norminfty}\\ 
		N_{\infty}:= \Varrho_1 (q-1)  &+ \Varrho_2 m   +\max \{ \Varrho_1, \Varrho_3     \}, \label{DefNinfty_Associated_With_Riemman} \\
		N_1:=\max \big\{\Varrho_1 (m+1), &\Varrho_2 (m+1), \Varrho_1 m + \Varrho_3 \big \},\label{DefNOne_Associated_With_Riemman}
	\end{align}
	where
	\begin{align}
		\Varrho_1 &:= 4 \underline \gamma_z^2 (1-\overline \rho_{zz} ) \frac{\zeta(s_{zz}+1)^2 }{s_{zz} e^{s_{zz}}} + 4  \overline \rho_{zz} \underline  \gamma_z ^2  \frac{ \zeta(2 s_z)  }{e^{s_{z}}},\\
		\Varrho_2 &:= 2 \Big(\overline \rho_{xz}   \big( \overline \gamma_x   \underline \gamma_z   \frac{ \zeta(2 s_z)  }{e^{s_{z}}} + \overline \gamma_z  \underline \gamma_x    \frac{ \zeta(2 s_x)}{ e^{s_x}}\big)  +\nonumber\\
		&\hspace{10mm} 2 (1-\overline \rho_{xz}) \underline \gamma_z  \underline \gamma_x \frac{ \zeta(s_{xz}+1)^2 }{ s_{xz}  e^{s_{xz}}}\Big) ,\\
		\Varrho_3 &:=  \frac{\Delta_{z}^2}{12} + \frac{\Delta_{z}^2   }{  \pi^{2 } e^{s_z} }  \zeta(2 s_z +2) + \frac{ 4 \overline\gamma_{z}^2  }{2^{s_z}   e^{s_z} }\zeta(2 s_z ),
	\end{align}
	with $\zeta(.)$ the Riemann zeta-function, $\sigma_{\text{min}}^Q$ is the smallest singular value of ${C^{x^Q \to z^Q }}(m,q)$, and 
	\begin{align}
		s_x = & \frac{2\pi^2 \underline \gamma_x^2}{\Delta_x^2}, \\
		s_z =& \frac{2\pi^2 \underline \gamma_z^2}{\Delta_z^2}, \\
		s_{xz} =& \frac{4\pi^2 \underline \gamma_x  \underline \gamma_z (1-\overline \rho_{xz} )}{\Delta_x \Delta_z},\\
		s_{zz} = & \frac{4\pi^2 \underline \gamma_z^2(1-\overline \rho_{zz} )}{\Delta_z^2}  .
	\end{align}  \hfill $\square$
\end{prop}
\begin{pf}
	See Appendix \ref{PropositionAssociatedWithReiman}.
\end{pf}
\begin{rem}
	Similar to Proposition \ref{NonUniformTheorem}, the RHS can be estimated from the quantized measurements, while the LHS can be determined {\em a priori} from the quantization parameters and the assumed bounds on the underlying signal statistics.
\end{rem}  
\begin{rem} 
	Similar to Remark \ref{SufficientCOnditionForAchiviebility_NonUniform}, we can derive the more stringent sufficient condition:
	\begin{align} \label{SuffCondBoundMid_tread_UniformUsingInftyNorm}
		\sqrt{N_{\infty} \hspace{1mm} N_1}< \frac{\sigma_{\text{min}}}{2 }, 
	\end{align}
	where the RHS no longer depends on the quantized signals. 
\end{rem}

\section{Bounds on quantization perturbation} \label{Upper_Bounds_On_Difference_Covariances_Mid_Tread} 
First let us obtain an upper bound on the difference between covariances of quantized and of unquantized signals. Suppose that zero-mean, jointly Gaussian scalar signals ${w_1}$ and ${w_2}$ are passed through  uniform quantizers with quantization intervals of length $\Delta_{w_1}$ and $\Delta_{w_2}$, respectively, with infinite number of quantization intervals.
The cross-covariances of uniformly quantized bivariate Gaussian vectors are derived in \cite{WidrowKollar}, in the form of infinite sums. \\
In Appendices \ref{Derivation_of_Eq_MainCovXXZeroMeanFinalMidTread1} and \ref{Derivation_of_Eq_MainCovXXZeroMeanFinalMidTread3}, it is shown that $|\gamma_{w^Q_1 w^Q_2} - \gamma_{w_1 w_2}| \leq  |\gamma_{w_1 \epsilon_2}| + |\gamma_{\epsilon_1 w_2}| + |\gamma_{\epsilon_1 \epsilon_2}|$ where
\begin{align} 
	|\gamma_{w_1 \epsilon_2}| &\leq \frac{|\rho_{w_1w_2}| \gamma_{w_1}  \gamma_{w_2}  k_2^{2s } s^{s} }{2^{s-1} \pi^{2s} e^{{s} }} \zeta(2s),\label{MainCovXXZeroMeanFinalMidTread2}\\ 
	|\gamma_{\epsilon_1 w_2}| &\leq \frac{|\rho_{w_1w_2}| \gamma_{w_1} \gamma_{w_2} k_1^{2\nu } \nu^{\nu}}{2^{\nu-1} \pi^{2\nu} e^{{\nu}}} \zeta(2\nu), \label{MainCovXXZeroMeanFinalMidTread1}\\
	|\gamma_{\epsilon_1 \epsilon_2}|&\leq \frac{(k_1k_2)^{\vartheta+1} \gamma_{w_1} \gamma_{w_2} \vartheta^{\vartheta} }{2^{2\vartheta} \pi^{2\vartheta+2} e^{\vartheta } (1 - |\rho_{w_1w_2}|)^\vartheta  } \zeta(\vartheta+1)^2,   \label{MainCovXXZeroMeanFinalMidTread3}
\end{align}
$k_i :=\frac{\Delta_{w_i}}{\gamma_{w_i}}, i=1,2$, $\gamma_{w_i}$ (with a single subscript) denotes  the standard deviation of  $w_i$, $\zeta(.)$ is the Riemann zeta-function, $\frac{1}{2} < s\leq \frac{2\pi^2 }{k_2^2},  \frac{1}{2} < \nu\leq \frac{2\pi^2 }{k_1^2}$, and $0< \vartheta \leq \frac{4\pi^2 (1-|\rho_{w_1 w_2}|)}{k_1 k_2}
$. Since  a-priori known ranges of standard deviations ($\underline \gamma_{w_i} \leq \gamma_{w_i} \leq \overline \gamma_{w_i}, i=1,2$) and of correlation coefficients are available, the intervals of $s$, $\nu$ and $\vartheta$ should be valid for  the entire ranges of the standard deviations and the correlation coefficients. Therefore, such variables for the upper bounds derived in \eqref{MainCovXXZeroMeanFinalMidTread2}-\eqref{MainCovXXZeroMeanFinalMidTread3} should be in the following intervals:
\begin{align} 
	\frac{1}{2} < s &\leq \frac{2\pi^2 }{K_1^2}, \label{RegionOfValidityFinal1} \\
	\frac{1}{2} <\nu &\leq \frac{2\pi^2 }{K_2^2} \label{RegionOfValidityFinal2},\\
	0 < \vartheta &\leq \frac{4\pi^2 (1-| \overline \rho |)}{K_1 K_2} \label{RegionOfValidityFinal3_1},
\end{align} 
where $K_i :=\frac{\Delta_{w_i}}{\underline \gamma_{w_i} }, i=1,2$.\\
In addition to cross-covariances between $x$ and $z$ and auto-covariances of $z$, the variance of $z$ is also present in the causality matrix. Therefore, to exploit Theorem \ref{GeneralTheoremForQuantization}, the difference between the variances of $z$ and $z^Q$ should also be derived as presented in Appendix \ref{PropositionAssociatedWithReiman_Variance}: 
\begin{align} \label{VarianceBoundForMidTread}
	|\gamma_{z^Q}^2 -\gamma_{z}^2| &\leq  \frac{\Delta_{z}^2}{12} + \frac{\Delta_{z}^2   }{  \pi^{2 } e^{\varsigma} }  \zeta(2 \varsigma +2) + \frac{ 4 \overline\gamma_{z}^2  }{2^\varsigma   e^{\varsigma} }\zeta(2 \varsigma ) , 
\end{align}
where $\varsigma :=\frac{2\pi^2  \underline \gamma_{z}^2}{\Delta_z^2}$.

\subsection{Derivation of equations (\ref{MainCovXXZeroMeanFinalMidTread2}) and (\ref{MainCovXXZeroMeanFinalMidTread1})} \label{Derivation_of_Eq_MainCovXXZeroMeanFinalMidTread1} 


First let us introduce an inequality for the exponential functions appearing in covariances of quantized signals as follows:
\begin{lem} \label{BoundOnExp}
	For real $0 < x \leq \frac{1}{s}, \ \ s \in \mathbb{R}_{>0}$:
	\begin{align} \label{InequalityFinal}
		e^{-\frac{1}{x}} \leq \frac{x^s s^s}{e^s}.
	\end{align}	
\end{lem}
\begin{pf}
	Using the monotonicity of the exponential function and the fact that $g(x) := -\frac{1}{x} - s \ln x + s-s\ln s, 0<x \leq \frac{1}{s}$ is always non-positive, the inequality \eqref{InequalityFinal} follows.
\end{pf} 
Moreover the following series is absolutely convergent:
\begin{align}\label{Application1}  
	\zeta(s+1) = \sum_{i=1}^{+\infty} \frac{1}{i^{s+1}},\ \  s\in \mathbb{R}_{>0}, 
\end{align}
where $\zeta(.)$ is the Riemann zeta-function.

Using \eqref{InequalityFinal}, \eqref{Application1} and comparison test theorem for series, it  can be shown that $\sum_{n_2=1}^{+\infty} e^{\frac{-2\pi^2n_2^2}{k_2^2}}$ is convergent and also $\sum_{n_2=1}^{+\infty} (-1)^{n_2} e^{\frac{-2\pi^2n_2^2}{k_2^2}}$ converges in the region where \eqref{InequalityFinal} holds. Exploiting the infinite series representing $\gamma_{w_1 \epsilon_2}$ derived in \cite{WidrowKollar}, Chapter 11,  the following bound can be obtained using \eqref{InequalityFinal} and \eqref{Application1}: 
\begin{align} \label{MainCovXEZeroMean1MidTread}
	|\gamma_{w_1 \epsilon_2}|&=  \Big| 2\rho_{w_1 w_2} \gamma_{w_1}  \gamma_{w_2}  \sum_{n_2=1}^{+\infty} (-1)^{n_2} e^{\frac{-2\pi^2n_2^2}{k_2^2}} \Big| \nonumber\\
	& \leq \frac{|\rho_{w_1 w_2}| \gamma_{w_1}  \gamma_{w_2}  k_2^{2s } {s}^{s}}{2^{s-1} \pi^{2s} e^{s} } \zeta(2s)
\end{align}
where $0 < s\leq \frac{2\pi^2 }{k_2^2}$. Furthermore, due to region of convergence of the Riemann zeta-function which is $2s>1$ in \eqref{MainCovXEZeroMean1MidTread}, we have:
\begin{align} \label{RegionOfValidity02}
	\frac{1}{2} < s\leq \frac{2\pi^2 }{k_2^2}.
\end{align}
And the region of quantization intervals should be as follows:
\begin{align} \label{RegionOfQuantizationInterval_Each}
	k_2< 2\pi, \ \ (\Delta_2 < 2\pi  \gamma_{w_2}).
\end{align}
to have \eqref{RegionOfValidity02}. Equation \eqref{MainCovXXZeroMeanFinalMidTread1} can be derived in a similar way.

\subsection{Derivation of equation (\ref{MainCovXXZeroMeanFinalMidTread3})} \label{Derivation_of_Eq_MainCovXXZeroMeanFinalMidTread3} 


Throughout the paper, when we deal with double infinite series,  convergence in Pringsheim sense \cite{bromwich1908introduction} is considered. 

Because of the absolute convergence of the Riemann zeta-function and Cauchy's multiplication theorem, the following holds: 
\begin{align}\label{Application2}  
	\lim\limits_{n_1,n_2 \to \infty} \sum_{i=1}^{n_1}\sum_{j=1}^{n_2} \frac{1}{i^{\vartheta+1}} \frac{1}{j^{\vartheta+1}} =  \zeta^2(\vartheta+1),
\end{align}
which implies that $\sum_{n_1=1}^{+\infty} \sum_{n_2=1}^{+\infty} \frac{(k_1k_2)^{\vartheta} \vartheta^\vartheta}{2^{2\vartheta} \pi^{2\vartheta} e^\vartheta  (1-\rho_{w_1 w_2})^\vartheta }  \frac{1}{n_1^{\vartheta+1}n_2^{\vartheta+1}}$ converges.
Furthermore:
\begin{align} \label{ReimannPopsUp}
	e^{-4\pi^2 \big (\frac{1}{2}(\frac{n_1 }{k_1})^2 +\frac{1}{2}(\frac{n_2 }{k_2})^2-\rho_{w_1 w_2}\frac{n_1 n_2}{k_1 k_2}\big)} &\leq   e^{\frac{-4\pi^2 (1-\rho_{w_1 w_2}) n_1 n_2 }{k_1 k_2}} \nonumber \\
	\leq  &\frac{(k_1k_2)^{\vartheta} \vartheta^\vartheta}{2^{2\vartheta} \pi^{2\vartheta} e^\vartheta  (1-\rho_{w_1 w_2})^\vartheta }  \frac{1}{n_1^{\vartheta}n_2^{\vartheta}}.
\end{align}
where the first inequality follows from Arithmetic Mean-Geometric Mean inequality and the second inequality is obtained by \eqref{InequalityFinal}. Thus, using  \eqref{ReimannPopsUp} and Comparison Test Theorem for double series \cite{bromwich1908introduction}, it can be shown that  $\sum_{n_1=1}^{+\infty} \sum_{n_2=1}^{+\infty}  \frac{1}{n_1n_2} e^{-4\pi^2 \big (\frac{1}{2}(\frac{n_1 }{k_1})^2 +\frac{1}{2}(\frac{n_2 }{k_2})^2-\rho_{w_1 w_2}\frac{n_1 n_2}{k_1 k_2}\big)}$ converges in the region  $0< k_1 k_2 \leq \frac{4\pi^2 (1-|\rho_{w_1 w_2}|)}{\vartheta}, \vartheta>0$. It can be shown that the following series is also convergent:
\begin{align}
	\sum_{n_1=1}^{+\infty} \sum_{n_2=1}^{+\infty}  \frac{(-1)^{n_1+n_2}}{n_1 n_2} e^{-4\pi^2 \big (\frac{1}{2}(\frac{n_1 }{k_1})^2 +\frac{1}{2}(\frac{n_2 }{k_2})^2-\rho_{w_1 w_2}\frac{n_1 n_2}{k_1 k_2}\big)}.
\end{align}
Using the order limit theorem for doubly index sequences, 
an upper bound can be derived as follows:
\begin{align} \label{ConvAndBound}
	\sum_{n_1=1}^{+\infty} \sum_{n_2=1}^{+\infty} \frac{(-1)^{n_1+n_2}}{n_1 n_2} e^{-4\pi^2 \big (\frac{1}{2}(\frac{n_1 }{k_1})^2 +\frac{1}{2}(\frac{n_2 }{k_2})^2-\frac{\rho_{w_1 w_2} n_1 n_2}{k_1 k_2}\big)} \leq  \nonumber \\
	\frac{(k_1k_2)^{\vartheta} \vartheta^\vartheta}{2^{2\vartheta} \pi^{2\vartheta} e^\vartheta (1-\rho_{w_1 w_2})^\vartheta } \sum_{n_1=1}^{+\infty} \sum_{n_2=1}^{+\infty} \frac{1}{n_1^{\vartheta+1}n_2^{\vartheta+1}}.
\end{align}
The covariance between quantization error terms $\epsilon_1$ and $\epsilon_2$ is expressed as follows in \cite{WidrowKollar}, Chapter 11: 
\begin{align}\label{MainCovEEZeroMean2MidTread}  
	\gamma_{\epsilon_1 \epsilon_2}= & \frac{k_1 k_2 \gamma_{w_1} \gamma_{w_2} }{2\pi^2}\sum_{n_1=1}^{+\infty} \sum_{n_2=1}^{+\infty}  \frac{(-1)^{n_1+n_2}}{n_1n_2} \times \nonumber\\ 
	& e^{-4\pi^2 \big (\frac{1}{2}(\frac{n_1 }{k_1})^2 +\frac{1}{2}(\frac{n_2 }{k_2})^2-\frac{\rho_{w_1 w_2} n_1 n_2}{k_1 k_2}\big)} \Big ( 1- e^{-4\pi^2 \big (\frac{2\rho_{w_1 w_2} n_1  n_2 }{k_1 k_2}\big) }   \Big).
\end{align}
For $0 < \rho_{w_1 w_2} < 1$, we have:
\begin{align}\label{MainCovEEZeroMean3MidTread}  
	|\gamma_{\epsilon_1 \epsilon_2}| &\leq  \frac{k_1 k_2 \gamma_{w_1}  \gamma_{w_2} }{2\pi^2} \sum_{n_1=1}^{+\infty} \sum_{n_2=1}^{+\infty}  \frac{1}{n_1n_2} \times \nonumber\\
	&\quad e^{-4\pi^2 \big (\frac{1}{2}(\frac{n_1 }{k_1})^2 +\frac{1}{2}(\frac{n_2 }{k_2})^2-\rho_{w_1 w_2}\frac{n_1 n_2}{k_1 k_2}\big)} \Bigg | 1- e^{-4\pi^2 \big (\frac{2\rho_{w_1 w_2} n_1  n_2 }{k_1 k_2}\big) }  \Bigg | \nonumber \\
	&\leq \frac{k_1 k_2 \gamma_{w_1}  \gamma_{w_2} }{2\pi^2} \sum_{n_1=1}^{+\infty} \sum_{n_2=1}^{+\infty}  \frac{2 e^{\frac{-2\pi^2 \big (n_1^2 k_2^2 + n_2^2 k_1^2 - 2\rho_{w_1 w_2} n_1 n_2 k_1 k_2\big) }{k_1^2 k_2^2}}}{n_1n_2} \nonumber\\
	&\leq\frac{(k_1k_2)^{\vartheta+1} \gamma_{w_1}  \gamma_{w_2} \vartheta^\vartheta }{2^{2\vartheta} \pi^{2\vartheta+2} e^\vartheta (1-\rho_{w_1 w_2})^\vartheta} \zeta(\vartheta+1)^2
\end{align}
where the last inequality follows from \eqref{Application2} and \eqref{ReimannPopsUp}.  In fact, it can be shown in a similar way to \eqref{MainCovEEZeroMean3MidTread} that:
\begin{align}\label{MainCovEEZeroMeanFinalMidTread}  
	|\gamma_{\epsilon_1 \epsilon_2}| &\leq \frac{(k_1k_2)^{\vartheta+1} \gamma_{w_1}  \gamma_{w_2} \vartheta^\vartheta}{2^{2\vartheta} \pi^{2\vartheta+2} e^\vartheta (1 - |\rho_{w_1 w_2}|)^\vartheta} \zeta(\vartheta+1)^2,\ \ |\rho_{w_1 w_2}|<1.
\end{align}
Note that $\vartheta$ should be chosen such that the following inequality holds for all positive integers $n_1$ and $n_2$:
\begin{align}\label{RegionOfValidity0}  
	\frac{k_1 k_2}{4\pi^2 (1-|\rho_{w_1 w_2}|) n_1 n_2 } \leq \frac{1}{\vartheta}
\end{align}
Hence, $\vartheta$ is as follows:
\begin{align}\label{RegionOfValidity1}  
	0< \vartheta \leq \frac{4\pi^2 (1-|\rho_{w_1 w_2}|)}{k_1 k_2}.
\end{align}
Note that the region of $\vartheta$ for convergence of the Riemann zeta-function is satisfied for any $\vartheta > 0$ in \eqref{MainCovEEZeroMeanFinalMidTread}.

\subsection{Derivation of equation \eqref{VarianceBoundForMidTread} }\label{PropositionAssociatedWithReiman_Variance} 
The variance of quantized signal $z^Q$ can be represented as follows \cite{WidrowKollar}, Chapter 11:
\begin{align}   
	\gamma_{z^Q}^2 = \gamma_{z}^2 + \frac{\Delta_{z}^2}{12} + \frac{\Delta_{z}^2}{\pi^2} \sum_{n=1}^{+\infty} \frac{(-1)^n}{n^2} e^{-\frac{2\pi^2 n^2}{k_z^2}}+ 
	4 \gamma_{z}^2  \sum_{n=1}^{+\infty} (-1)^n e^{-\frac{2\pi^2 n^2}{k_z^2}}.
\end{align}	
Using \eqref{InequalityFinal}, we have:
\begin{align}
	|\gamma_{z^Q}^2 -\gamma_{z}^2| &\leq \frac{\Delta_{z}^2}{12} + \frac{\Delta_{z}^2}{\pi^2} \sum_{n=1}^{+\infty} \frac{1}{n^2} e^{-\frac{2\pi^2 n^2}{k_z^2}}+ 
	4 \gamma_{z}^2  \sum_{n=1}^{+\infty}  e^{-\frac{2\pi^2 n^2}{k_z^2}} \nonumber\\
	&\leq \frac{\Delta_{z}^2}{12} + \frac{\Delta_{z}^2 k_z^{2 \varsigma	} \varsigma^\varsigma }{2^\varsigma \pi^{2(\varsigma +1)} e^{\varsigma} }  \zeta(2 \varsigma +2) + \frac{ 4\gamma_{z}^2 k_z^{2 \varsigma	} \varsigma^\varsigma }{2^\varsigma \pi^{2\varsigma } e^{\varsigma} }\zeta(2 \varsigma ) \nonumber\\
	&\leq \frac{\Delta_{z}^2}{12} + \frac{\Delta_{z}^2 k_z^{2 \varsigma	} \varsigma^\varsigma }{2^\varsigma \pi^{2(\varsigma +1)} e^{\varsigma} }  \zeta(2 \varsigma +2) + \frac{ 4 \overline\gamma_{z}^2 k_z^{2 \varsigma	} \varsigma^\varsigma }{2^\varsigma \pi^{2\varsigma } e^{\varsigma} }\zeta(2 \varsigma ) \label{BoundOnVariance_WithReiman}
\end{align}
where $	\frac{1}{2} < \varsigma \leq \frac{2\pi^2 }{k_z^2}$ due to Lemma \ref{BoundOnExp} and the convergence region of the Riemann zeta-function. Furthermore, since $k_z =\frac{\Delta_{z}}{ \gamma_{z} }$ and a-priori known range of $\gamma_{z}$, i.e.   $ \underline \gamma_{z}  \leq \gamma_{z} \leq \overline \gamma_{z} $, is known,  $\varsigma$ should be in the region of $	\frac{1}{2} < \varsigma \leq \frac{2\pi^2  \underline \gamma_{z}^2}{\Delta_z^2}$. Note that the value of $\varsigma$ in \eqref{BoundOnVariance_WithReiman}  should be chosen   such that the upper bound on $|\gamma_{z^Q}^2 -\gamma_{z}^2|$ is minimized. It can be shown that the second and third terms in \eqref{BoundOnVariance_WithReiman} are monotonically decreasing with respect to $\varsigma$. Thus, $\varsigma$ should be chosen to equal $\frac{2\pi^2  \underline \gamma_{z}^2}{\Delta_z^2}$ in order to have a tighter bound. Substituting $\varsigma=\frac{2\pi^2  \underline \gamma_{z}^2}{\Delta_z^2}$ in \eqref{BoundOnVariance_WithReiman}, it can be shown that:
\begin{align}
	|\gamma_{z^Q}^2 -\gamma_{z}^2| &\leq  \frac{\Delta_{z}^2}{12} + \frac{\Delta_{z}^2   }{  \pi^{2 } e^{\varsigma} }  \zeta(2 \varsigma +2) + \frac{ 4 \overline\gamma_{z}^2  }{2^\varsigma   e^{\varsigma} }\zeta(2 \varsigma ) , \label{BoundOnVariance_WithReiman_Final}
\end{align} 
which is \eqref{VarianceBoundForMidTread}.     

\section{Proof of Proposition \ref{MidTreadProposition}}\label{PropositionAssociatedWithReiman}
In order to develop a bound on the norm of quantization perturbation ($\| {\Gamma_{\epsilon}}(m,q)  \|$) to exploit Theorem \ref{GeneralTheoremForQuantization} for the infinite-level quantized signals, the difference between elements of the causality matrix $C_G^{x \to z}(m,q)$  and $C^{x^Q \to z^Q}(m,q)$  should be first obtained. 
The elements of the causality matrix   $C_G^{x \to z}(m,q)$ are  the cross-covariances between $x$ and $z$ and  the auto-covariances of signal $z$. 
The infinity norm of ${\Gamma_{\epsilon}} (m,q) $ is as follows:
\begin{align} \label{BoundForInfty_MidUnif}
	\| {\Gamma_{\epsilon}} (m,q) \|_\infty  =& \| C^{x^Q \to z^Q}(m,q) - C_G^{x \to z}(m,q) \|_\infty \nonumber\\
	=& \max_{1\leq i\leq m+1} \sum_{j=1}^{m+q} |\gamma_{\epsilon_{ij}}|\nonumber\\
	=&\max \Big\{ \sum_{j=1}^{m+q} |\gamma_{\epsilon_{1j}}|, \max_{2\leq i\leq m+1} \sum_{j=1}^{m+q} |\gamma_{\epsilon_{ij}}|\Big \} \nonumber\\
	\leq& m \sup_{\kappa} |\gamma_{x^Qz^Q} (\kappa) - \gamma_{xz} (\kappa)| +  \nonumber\\
	&(q-1) \sup_{\kappa \neq 0} |\gamma_{z^Qz^Q} (\kappa) - \gamma_{zz} (\kappa)| + \nonumber\\
	&\max \big\{   \sup_{\kappa \neq 0} |\gamma_{z^Qz^Q} (\kappa) - \gamma_{zz} (\kappa)|,     |\gamma_{z^Q}^2 - \gamma_{z}^2 |      \big \}.
\end{align} 
An upper bound on  $\| {\Gamma_{\epsilon}} (m,q) \|_\infty$ can be derived as follows:
\begin{align} \label{BoundForInfty3}
	\| {\Gamma_{\epsilon}} (m,q) \|_\infty \leq& m I_{xz}+  (q-1) I_{zz}    + \max \Big\{ I_{zz}, |\gamma_{z^Q}^2 - \gamma_{z}^2 |     \Big \}
\end{align}	
where $I_{xz}:= \sup_{\kappa} |\gamma_{x^Qz^Q} (\kappa) - \gamma_{xz} (\kappa)|$ and $I_{zz}:= \sup_{\kappa \neq 0} |\gamma_{z^Qz^Q} \allowbreak (\kappa) - \gamma_{zz} (\kappa)|$ are obtained as follows by \eqref{MainCovXXZeroMeanFinalMidTread2}-\eqref{MainCovXXZeroMeanFinalMidTread3} and by replacing the standard deviations and correlation coefficients with their bounds:
\begin{align}
	&I_{xz} = f_1( \overline \rho_{xz} \overline \gamma_z , \underline \gamma_x , \Delta_x,s_x)+ f_1(\overline \rho_{xz} \overline \gamma_x , \underline \gamma_z , \Delta_z,s_z) + \nonumber\\
	&\hspace{9mm} f_2(\underline \gamma_z \underline \gamma_x  (1-\overline \rho_{xz} ),\Delta_x \Delta_z, s_{xz}), \label{DefIxz}\\
	&I_{zz} := 2 f_1(\overline \rho_{zz}\underline \gamma_z^{-1} ,\underline \gamma_z, \Delta_z ,s_z) + f_2\big( \underline \gamma_z^2 (1-\overline \rho_{zz} ),\Delta_z^2,s_{zz}\big) , \label{DefIzz}\\
	&f_1(a,b,c,d) := \frac{a c^{2d} d^d}{2^{d-1}\pi^{2d} e^{d} b^{2d-1}}\zeta(2d), \\
	&f_2(a,b,c) := \frac{b^{c+1} c^c}{2^{2c}\pi^{2c+2} e^{c}a^c}\zeta(c+1)^2 \label{Deff2},
\end{align}
and 
\begin{align} \label{DefRhomax}
	\overline\rho_{xz} := \max_{\kappa } \big \{|\rho_{xz}(\kappa)| \big \},\\
	\overline \rho_{zz}  := \max_{\kappa \neq 0} \big \{|\rho_{zz}(\kappa)| \big \}.
\end{align}
A bound on the norm 1 can be also obtained as follows:
\begin{align} \label{BoundForOneNorm}
	\|{\Gamma_{\epsilon}} (m,q) \|_1 :=& \| C^{x^Q \to z^Q}(m,q) - C_G^{x \to z}(m,q) \|_1 \nonumber\\
	\leq &  \max \Big\{(m+1) \sup_{\kappa} |\gamma_{x^Qz^Q} (\kappa) - \gamma_{xz} (\kappa)|, \nonumber\\
	& \hspace{9mm}(m+1) \sup_{\kappa \neq 0} |\gamma_{z^Qz^Q} (\kappa) - \gamma_{zz} (\kappa)|,          \nonumber\\
	& \hspace{9mm} m \sup_{\kappa \neq 0} |\gamma_{z^Qz^Q} (\kappa) - \gamma_{zz} (\kappa)|  +  |\gamma_{z^Q}^2  - \gamma_{z}^2 | \Big \} \nonumber \\ 
	\leq &  \max \Big\{(m+1) I_{xz}, (m+1) I_{zz}, m I_{zz}+     |\gamma_{z^Q}^2  - \gamma_{z}^2 |  \Big \}.
\end{align}	
The term $|\gamma_{z^Q}^2  - \gamma_{z}^2 |$ appearing in \eqref{BoundForInfty3} and \eqref{BoundForOneNorm}  should be replaced by its upper bound derived in \eqref{VarianceBoundForMidTread}. 
Moreover, note that $I_{xz}$ and $I_{zz}$ appearing in the norm one and infinity of ${\Gamma_{\epsilon}} (m,q)$ depend on $s_{x}$, $s_{z}$, $s_{xz}$ and $s_{zz}$. Such variables should be chosen such that the upper bound on $\|{\Gamma_{\epsilon}} (m,q)\|_2$  ($\|{\Gamma_{\epsilon}} (m,q)\|_2 \leq \sqrt{  \|{\Gamma_{\epsilon}} (m,q)\|_1  \|{\Gamma_{\epsilon}} (m,q)\|_\infty}$) is minimized and be also in the intervals defined in \eqref{RegionOfValidityFinal1}-\eqref{RegionOfValidityFinal3_1}. It  can be shown that the functions $f_1$ and $f_2$ are monotonically decreasing with respect to such variables. Thus, it is sufficient to evaluate the upper bound on  $\|{\Gamma_{\epsilon}} (m,q)\|_2$  just for the values of $s_x$, $s_z$, $s_{xz}$ and $s_{zz}$ mentioned in the following:
\begin{align}
	s_x = & \frac{2\pi^2 \underline \gamma_x^2}{\Delta_x^2},\label{OptimalSx}\\
	s_z =& \frac{2\pi^2 \underline \gamma_z^2}{\Delta_z^2},\label{OptimalSz}\\
	s_{xz} =& \frac{4\pi^2 \underline \gamma_x  \underline \gamma_z (1-\overline \rho_{xz} )}{\Delta_x \Delta_z},\\
	s_{zz} = & \frac{4\pi^2 \underline \gamma_z^2(1-\overline \rho_{zz} )}{\Delta_z^2} \label{OptimalSzz},
\end{align}
where 	$\overline\rho_{xz} := \max_{\kappa } \big \{|\rho_{xz}(\kappa)| \big \}$ and $\overline \rho_{zz}  := \max_{\kappa \neq 0} \big \{|\rho_{zz}(\kappa)| \big \}$. Substituting such optimal values in \eqref{DefIxz} and \eqref{DefIzz} yields \eqref{DefNinfty_Associated_With_Riemman} and \eqref{DefNOne_Associated_With_Riemman} and Theorem \ref{GeneralTheoremForQuantization} implies Proposition \ref{MidTreadProposition}.  Note that the value of $\varsigma$ in \eqref{VarianceBoundForMidTread} equals to $s_z$ in \eqref{OptimalSz}. Therefore, for ease of notation, $\varsigma$ is replaced by $s_z$ in the Proposition.

\section{Derivations of equations (\ref{MainCovXEZeroMean7}) and (\ref{MainCovEXZeroMean7})} \label{Big_Oh_Derivations_EW}
Using the comparison test theorem for infinite series, it can be shown that $\sum_{n_2=1}^{+\infty} (-1)^{n_2} e^{-\frac{2\pi^2n_2^2}{k_2^2}}$ is absolutely convergent.  The expression of $\gamma_{w_1 \epsilon_2}$ is introduced in \cite{WidrowKollar}, Chapter 11. The series can be bounded as follows:
\begin{align} \label{MainCovXEZeroMean1MidTread_HR}
	|\gamma_{w_1 \epsilon_2}|&=  \Big| 2\rho_{w_1 w_2} \gamma_{w_1}  \gamma_{w_2}  \sum_{n_2=1}^{+\infty} (-1)^{n_2} e^{-\frac{2\pi^2n_2^2}{k_2^2}} \Big| \nonumber\\
	& <  2  |\rho_{w_1 w_2} | \gamma_{w_1}  \gamma_{w_2} \sum_{n_2=1}^{+\infty} e^{-\frac{2\pi^2n_2^2}{k_2^2}}\nonumber\\
	& < 2  |\rho_{w_1 w_2} | \gamma_{w_1}  \gamma_{w_2} \sum_{n_2=1}^{+\infty} e^{-\frac{2\pi^2n_2 }{k_2^2}}\nonumber\\
	& < 2  |\rho_{w_1 w_2} | \gamma_{w_1}  \gamma_{w_2} \frac{e^{-\frac{2\pi^2 }{k_2^2}}}{1- e^{-\frac{2\pi^2 }{k_2^2}}}.
\end{align}
For sufficiently fine $k_2$, it can be written as:
\begin{align} \label{MainCovXEZeroMean1MidTread_HR_Final}
	|\gamma_{w_1 \epsilon_2}| < 4  |\rho_{w_1 w_2} | \gamma_{w_1}  \gamma_{w_2} e^{-\frac{2\pi^2 }{k_2^2}} = 4  |\gamma_{w_1 w_2} |   e^{-\frac{2\pi^2 }{k_2^2}},
\end{align}
and \eqref{MainCovXEZeroMean7} follows.  The same method can be used to derive \eqref{MainCovEXZeroMean7}.
\section{Derivation of equation (\ref{MainCovEEZeroMean7})} \label{Big_Oh_Derivation_EE}
The covariance between quantization error terms $\epsilon_1$ and $\epsilon_2$ is as follows \cite{WidrowKollar}, Chapter 11:
\begin{align} 
	\gamma_{\epsilon_1 \epsilon_2}= & \frac{k_1 k_2 \gamma_{w_1} \gamma_{w_2} }{2\pi^2}\sum_{n_1=1}^{+\infty} \sum_{n_2=1}^{+\infty}  \frac{(-1)^{n_1+n_2}}{n_1n_2} \times \nonumber\\ 
	& e^{-4\pi^2 \big (\frac{1}{2}(\frac{n_1 }{k_1})^2 +\frac{1}{2}(\frac{n_2 }{k_2})^2-\frac{\rho_{w_1 w_2} n_1 n_2}{k_1 k_2}\big)} \Big ( 1- e^{-4\pi^2 \big (\frac{2\rho_{w_1 w_2} n_1  n_2 }{k_1 k_2}\big) }   \Big).
\end{align}
It can be shown that it is convergent and a bound as follows can be obtained:
\begin{align}\label{MainCovEEZeroMean3MidTread_HR}  
	|\gamma_{\epsilon_1 \epsilon_2}| &<  \frac{k_1 k_2 \gamma_{w_1}  \gamma_{w_2} }{2\pi^2} \sum_{n_1=1}^{+\infty} \sum_{n_2=1}^{+\infty}  \frac{1}{n_1n_2}  \Bigg | 1- e^{-4\pi^2 \big (\frac{2 |\rho_{w_1 w_2}| n_1  n_2 }{k_1 k_2}\big) }  \Bigg | \times \nonumber\\
	&\quad e^{-4\pi^2 \big (\frac{1}{2}(\frac{n_1 }{k_1})^2 +\frac{1}{2}(\frac{n_2 }{k_2})^2-|\rho_{w_1 w_2}|\frac{n_1 n_2}{k_1 k_2}\big)}  \nonumber \\
	&< \frac{k_1 k_2 \gamma_{w_1}  \gamma_{w_2} }{ \pi^2} \sum_{n_1=1}^{+\infty} \sum_{n_2=1}^{+\infty}   e^{-2\pi^2 \big ( (\frac{n_1 }{k_1})^2 + (\frac{n_2 }{k_2})^2-2|\rho_{w_1 w_2}|\frac{n_1 n_2}{k_1 k_2}\big)} \nonumber\\
	&<  \frac{k_1 k_2 \gamma_{w_1}  \gamma_{w_2} }{ \pi^2} \sum_{n_1=1}^{+\infty} \sum_{n_2=1}^{+\infty}   e^{-2\pi^2   (1- |\rho_{w_1 w_2}| )\big( (\frac{n_1 }{k_1})^2 + (\frac{n_2}{k_2})^2 \big)},
\end{align}
where the last inequality follows from Rayleigh quotient theorem. And
\begin{align}
	|\gamma_{\epsilon_1 \epsilon_2}| &<    \frac{k_1 k_2 \gamma_{w_1}  \gamma_{w_2} }{ \pi^2} \sum_{n_1=1}^{+\infty} \sum_{n_2=1}^{+\infty}   e^{-2\pi^2   (1- |\rho_{w_1 w_2}|)\big( \frac{n_1 }{k_1^2}  + \frac{n_2}{k_2^2} \big)}.
\end{align}
Note that since $\sum_{n =1}^{+\infty}   e^{- \frac{2\pi^2  (1- |\rho_{w_1 w_2}|)}{k_i^2} n }, i=1,2$ are convergent for $|\rho_{w_1 w_2}|\neq 1$, it can be shown that:
\begin{align} \label{Epsilon1Epsilon2_HR}
	|\gamma_{\epsilon_1 \epsilon_2}| &<    \frac{k_1 k_2 \gamma_{w_1}  \gamma_{w_2} }{ \pi^2} \frac{ e^{- \frac{2\pi^2  (1- |\rho_{w_1 w_2}|)}{k_1^2}  }}{1- e^{- \frac{2\pi^2  (1- |\rho_{w_1 w_2}|)}{k_1^2}  }}\frac{ e^{- \frac{2\pi^2  (1- |\rho_{w_1 w_2}|)}{k_2^2}  }}{1- e^{- \frac{2\pi^2  (1- |\rho_{w_1 w_2}|)}{k_2^2}  }}.
\end{align} 
For sufficiently fine $k_1$ and $k_2$, we have:
\begin{align}
	\gamma_{\epsilon_1 \epsilon_2} = O\big( k_1 k_2 e^{-2\pi^2  (1- |\rho_{w_1 w_2}|) (\frac{1}{k_1^2} + \frac{1}{k_2^2})} \big).
\end{align}
\section{Derivation of equation (\ref{MainVar4})} \label{Big_Oh_Derivations_Var}
The relation between variance of quantized and unquantized signals is as follows \cite{WidrowKollar}, Chapter 11:
\begin{align}   
	\gamma_{w_1^Q}^2 =& \gamma_{w_1}^2 + \frac{k_1^2 \gamma_{w_1}^2 }{12} + \frac{\Delta_{w_1}^2}{\pi^2} \sum_{n=1}^{+\infty} \frac{(-1)^n}{n^2} e^{-\frac{2\pi^2 n^2}{k_1^2}}+ \nonumber\\
	& 4 \gamma_{w_1}^2  \sum_{n=1}^{+\infty} (-1)^n e^{-\frac{2\pi^2 n^2}{k_1^2}}.
\end{align}
Therefore, we have:
\begin{align}   \label{Var_HR}
	\bigg|\gamma_{w_1^Q}^2 - \gamma_{w_1}^2 - \frac{k_1^2 \gamma_{w_1}^2 }{12} \bigg| <&  \frac{\Delta_{w_1}^2}{\pi^2} \sum_{n=1}^{+\infty} \frac{1}{n^2} e^{-\frac{2\pi^2 n^2}{k_1^2}} + 4 \gamma_{w_1}^2  \sum_{n=1}^{+\infty}  e^{-\frac{2\pi^2 n^2}{k_1^2}} \nonumber \\
	<&  (\frac{\Delta_{w_1}^2}{\pi^2} + 4 \gamma_{w_1}^2 ) \sum_{n=1}^{+\infty}  e^{-\frac{2\pi^2 n^2}{k_1^2}} \nonumber\\
	<& (\frac{\Delta_{w_1}^2}{\pi^2} + 4 \gamma_{w_1}^2 ) \frac{e^{-\frac{2\pi^2 }{k_1^2}}}{1- e^{-\frac{2\pi^2 }{k_1^2}}},
\end{align}
and  for sufficiently small $k_1$, \eqref{MainVar4} follows.
\section{Upper bound on $\| {\Gamma_{\epsilon}} (m,q) \|_2$  in high-resolution regime } \label{UpperBound_NormTwo_HR}
Norm infinity of the matrix  ${\Gamma_{\epsilon}} (m,q)$ can be written as follows in high resolution regime:
\begin{align} \label{BoundForInfty_HR}
	\| {\Gamma_{\epsilon}} (m,q) \|_\infty  :=& \max_{1\leq i\leq m+1} \sum_{j=1}^{m+q} |\gamma_{\epsilon_{ij}}|\nonumber\\
	\leq&  m \sup_{\kappa} |\gamma_{x^Qz^Q} (\kappa) - \gamma_{xz} (\kappa)| + (q-1) \times\nonumber \\ 
	& \sup_{\kappa \neq 0} |\gamma_{z^Qz^Q} (\kappa) - \gamma_{zz} (\kappa)|+ \nonumber\\
	& \max \big\{ \sup_{\kappa \neq 0} |\gamma_{z^Qz^Q} (\kappa) - \gamma_{zz} (\kappa)| , |\gamma_{z^Q}^2 - \gamma_{z}^2| \big\},    
\end{align}
where
\begin{align}
	\sup_{\kappa} |\gamma_{x^Qz^Q} (\kappa) - \gamma_{xz} (\kappa)| &<    4  \overline  \gamma_{xz}   e^{-\frac{2\pi^2}{k_x^2}}    +  O\big(  e^{-\frac{2\pi^2}{k_z^2}}  + \nonumber\\
	& \hspace{5mm} k_x k_z e^{-2\pi^2  (1- \overline \rho_{xz}) (\frac{1}{k_x^2} + \frac{1}{k_z^2})} \big), \label{S_xz_HR}
	\\
	\sup_{\kappa \neq 0} |\gamma_{z^Qz^Q} (\kappa) - \gamma_{zz} (\kappa)| &= O\big( e^{-\frac{2\pi^2}{k_z^2}} +  k_z^2 e^{-\frac{4\pi^2}{k_z^2}  (1-  \overline \rho_{zz} )  } \big),
	\\
	|\gamma_{z^Q}^2 - \gamma_{z}^2| &= \frac{k_z^2 \gamma_{z}^2 (=\Delta_{z}^2)}{12} + O (   e^{-\frac{ 2\pi^2}{k_z^2}}),
\end{align}
are obtained from \eqref{MainCovXEZeroMean1MidTread_HR_Final} and \eqref{MainCovXEZeroMean7}-\eqref{MainVar4} and $\overline \gamma_{xz}$ is the upper bound on the cross-covariance between signals $x$ and $z$.

For sufficiently fine $k_z$, $\frac{k_z^2 \gamma_{z}^2}{12} \gg  e^{-\frac{2\pi^2}{k_z^2}}$ and  $\frac{k_z^2 \gamma_{z}^2}{12} \gg  k_z^2 e^{-\frac{4\pi^2}{k_z^2}  (1-  \rho_{zz} ^{\max})  }$. Hence, the dominant term between terms appearing in $\sup_{\kappa \neq 0} |\gamma_{z^Qz^Q} (\kappa) - \gamma_{zz} (\kappa)|$ and $|\gamma_{z^Q}^2 - \gamma_{z}^2|$ is $\frac{k_z^2 \gamma_{z}^2}{12}$. 

Furthermore, $\frac{k_z^2 \gamma_{z}^2}{12} \gg  k_x k_z e^{-2\pi^2  (1-  \rho_{xz} ^{\max}) (\frac{1}{k_x^2} + \frac{1}{k_z^2})}$ since in high-resolution regime $\frac{k_z  \gamma_{z}^2}{12} e^{ \frac{2\pi^2}{k_z^2}   (1- \rho_{xz}^{\max})  } \gg  k_x   e^{- \frac{2\pi^2}{k_x^2}   (1-  \rho_{xz}^{\max}) }$.  Therefore, the following can be stated:
\begin{align}\label{Norm_Infinity_HR}
	\| {\Gamma_{\epsilon}} (m,q) \|_\infty \leq  	4  m \overline   \gamma_{xz}  e^{-\frac{2\pi^2}{k_x^2}}+ \frac{k_z^2 \gamma_{z}^2}{12}
\end{align} 
Based on what is mentioned above, the norm one $\| {\Gamma_{\epsilon}} (m,q) \|_1$ can be written as follows:
\begin{align}\label{Norm_One_HR}
	\| {\Gamma_{\epsilon}} (m,q) \|_1 \leq& \max \big \{(m+1)\sup_{\kappa} |\gamma_{x^Qz^Q} (\kappa) - \gamma_{xz} (\kappa)| ,\nonumber\\
	&\hspace{10mm} m \sup_{\kappa \neq 0} |\gamma_{z^Qz^Q} (\kappa) - \gamma_{zz} (\kappa)| +  |\gamma_{z^Q}^2 - \gamma_{z}^2|  \big \}\nonumber\\
	\leq&\max\Big\{4(m+1) \overline   \gamma_{xz}  e^{-\frac{2\pi^2}{k_x^2}} , \frac{k_z^2 \gamma_{z}^2}{12} \Big\}
\end{align}
Using \eqref{Norm_Infinity_HR} and \eqref{Norm_One_HR}:
\begin{align} \label{Norm_Two_HR_1}
	\| {\Gamma_{\epsilon}} (m,q) \|_2 \leq &   \Big( \big(4 m  \overline   \gamma_{xz}   e^{-\frac{2\pi^2}{k_x^2}}+ \frac{\Delta_{z}^2}{12}\big) \times \nonumber\\
	&\hspace{4mm}\max\Big\{ 4(m+1) \overline   \gamma_{xz}   e^{-\frac{2\pi^2}{k_x^2}}, \frac{\Delta_{z}^2}{12} \Big\} \Big)^\frac{1}{2} +\nonumber\\
	& \Big(4(m+1) \overline   \gamma_{xz}  e^{-\frac{2\pi^2}{k_x^2}}
	+ \frac{k_z^2 \gamma_{z}^2}{12} \Big)^{\frac{1}{2}} \times \nonumber\\
	& O\bigg(e^{-\frac{ \pi^2}{k_z^2}}  + \max\Big\{ ( k_x k_z)^{\frac{1}{2}} e^{- \pi^2  (1- \overline \rho_{xz}) (\frac{1}{k_x^2} + \frac{1}{k_z^2})} 	 , \nonumber\\
	&\hspace{26mm}   k_z e^{-\frac{2\pi^2}{k_z^2}  (1-  \overline \rho_{zz} )  } \Big\} \bigg) \nonumber\\
	\leq & 4 (m+1) \overline   \gamma_{xz}   e^{-\frac{2\pi^2}{k_x^2}} + \frac{\Delta_{z}^2}{12} +\nonumber\\
	& \Big(4(m+1) \overline   \gamma_{xz}  e^{-\frac{2\pi^2}{k_x^2}}
	+ \frac{\Delta_{z}^2}{12} \Big)^{\frac{1}{2}} \times \nonumber\\
	& O\bigg(e^{-\frac{ \pi^2}{k_z^2}}  + \max\Big\{ ( k_x k_z)^{\frac{1}{2}} e^{- \pi^2  (1- \overline \rho_{xz}) (\frac{1}{k_x^2} + \frac{1}{k_z^2})} 	 ,   \nonumber\\
	&\hspace{26mm} k_z  e^{-\frac{2\pi^2}{k_z^2}  (1-  \overline \rho_{zz} )  } \Big\} \bigg)
\end{align}  

which is \eqref{Norm_Two_HR_Final}.

\section{Proof of Theorem \ref{ErgodicityPreservation}} \label{Proof_Of_ErgodicityPreservation}
Let us first mention some definitions and make the definitions of their vector counterparts. \\
\begin{defn} [Stationary Process] \cite{Breiman}  
	A  process  $(y_k)_{k \geq 1}$  is  said  to  be  stationary  if $y_{k+1}, y_{k+2}, ... $  has  the  same  distribution  as $y_1, y_2, ...$   for 
	every  $k  \geq  1$;  that  is,  if  for  each $k  \geq  1$:
	\begin{align}
		P\big( (y_1, y_2, ... ) \in B \big) = P\big( (y_{k+1}, y_{k+2}, ... ) \in B\big) 
	\end{align}
	for  every  $B \in \mathcal{B}_\infty$, where 
	the Borel field $\mathcal{B}_\infty$
	is the smallest $\sigma$-field of subsets of $\mathbb{R}^{\infty}$ containing all finite-dimensional rectangles, $\mathbb{R}^{\infty}$ denotes the space consisting of all infinite sequences $(y_1, y_2, ...)$	of real numbers, and an $n$-dimensional rectangle in $\mathbb{R}^{\infty}$ is a set of the form 
	\begin{align}
		\{{y} \in \mathbb{R}^{\infty}; y_1 \in I_1,..., y_n \in I_n\},
	\end{align} 
	where $I_1,..., I_n$ are finite or infinite intervals. 
\end{defn}
\begin{defn} [Invariant Set] \cite{Breiman}  
	An event $A$ is invariant if there exists $B \in  \mathcal{B}_\infty$ such that for every $k \geq 1$:
	\begin{align}
		A= \{(y_{k}, y_{k+1},...) \in B \}.
	\end{align}
\end{defn}
\begin{defn} [Ergodic Process] \cite{Breiman}  
	A stationary process   $(y_k)_{k \geq 1}$   is  ergodic  if 
	every  invariant  event  has  probability  zero  or  one. 
\end{defn}
The following can be defined for vector processes:
\begin{defn} [Stationary Vector Process] \label{Stationary_Vector}
	An  $l$-variate vector process  $(y_k)_{k \geq 1}$, where $y_k :=[y_{1,k},..., y_{l,k}]^\intercal$,  is  said  to  be  stationary  if ${y}_{k+1}, {y}_{k+2}, ... $  has  the  same  distribution  as $y_1, y_2, ...$   for 
	every  $k  \geq  1$;  that  is,  if  for  each $k  \geq  1$:
	\begin{align}
		P\big( ({y}_1, {y}_2, ... ) \in B \big) = P\big( ({y}_{k+1}, {y}_{k+2}, ... ) \in B\big) 
	\end{align}
	for  every  $B \in \mathbfcal{B}_\infty$, where $\mathbfcal{B}_\infty$ is the corresponding Borel field. 
\end{defn}
\begin{defn} [Invariant Set for Vector Processes]  
	An event $A$ is invariant if there exists $B \in \mathbfcal{B}_\infty$ such that for every $k \geq 1$:
	\begin{align}
		A= \{({y}_{k}, {y}_{k+1},...) \in B \}.
	\end{align}
\end{defn}
\begin{defn} [Ergodic Vector Processes]
	A stationary vector process  $(y_k)_{k \geq 1}$  is  ergodic  if 
	every  invariant  event  has  probability  zero  or  one. 
\end{defn}
Now let us return to the proof of Theorem \ref{ErgodicityPreservation}. 	According to Definition \ref{Stationary_Vector}, we need to show that for every $B_{\xi} \in \mathbfcal{B}_\infty$ and $k \geq 1$, the following holds:
\begin{align}
	&P\big(\omega: (\xi_1 (\omega), \xi_2 (\omega), ...) \in B_{\xi} \big)= \nonumber\\
	& P\big(\omega: (\xi_{k+1} (\omega), \xi_{k+2} (\omega),...) \in B_{\xi} \big).
\end{align}
We know that
\begin{align}
	&P\big(\omega: (\xi_1 (\omega), \xi_2 (\omega), ...) \in B_{\xi} \big) = \nonumber\\
	& P\bigg(\omega: \Big({f} \big({y}_1(\omega), {y}_{2}(\omega),...\big) , {f}\big({y}_2(\omega), {y}_{3}(\omega),...\big), ...\Big) \in B_{\xi} \bigg). 
\end{align}
Since ${f}$  is  a 
measurable  function, for any $B_{\xi}$ there exists $B_{y} $ such that:
\begin{align}
	&\Big\{\omega: \Big({f} \big({y}_1(\omega), {y}_{2}(\omega),...\big) , {f}\big({y}_2(\omega), {y}_{3}(\omega),...\big), ... \Big) \in B_{\xi} \Big \} \nonumber \\
	&= \Big\{\omega: \big({y}_1(\omega), {y}_{2}(\omega),...\big)  \in B_{y} \Big \} \label{Stationary1}
\end{align}
and
\begin{align} \label{Stationary2}
	\Big\{\omega: \Big(&{f} \big({y}_{k+1}(\omega), {y}_{k+2}(\omega),...\big) , {f}\big({y}_{k+2}(\omega), {y}_{k+3}(\omega),...\big), \nonumber\\
	& ...\Big) \in B_{\xi} \Big \} = \Big\{\omega: \big({y}_{k+1}(\omega), {y}_{k+2}(\omega),...\big)  \in B_{y} \Big \}.
\end{align}
Since  $(y_k)_{k \geq 1}$ is stationary due to the assumption of the Theorem, the probability of the RHS's of \eqref{Stationary1} and \eqref{Stationary2} is equal. Thus, the probability of the LHS's of \eqref{Stationary1} and \eqref{Stationary2} is equal for any $B_{\xi} \in \mathbfcal{B}_\infty$ and $k \geq 1$. It implies that  $(\xi_k)_{k \geq 1}$  is  a stationary vector process.\\
For ergodicity part of the Theorem, let us assume that $A$ is an invariant set for	$(\xi_k)_{k \geq 1}$. Therefore, there exists $ B^e_{\xi} \in \mathbfcal{B}_\infty$ such that for every $k \geq 1$:
\begin{align}\label{Ergodicity2}
	A = \big \{ \omega: (\xi_k(\omega),  \xi_{k+1}(\omega),  ...)  \in  B^e_{\xi} \big\}.
\end{align}
Since ${f}$  is  a 
measurable  function,  there exists $B^e_{y}$ for $B^e_{\xi}$ such that \eqref{Ergodicity2} can be written as follows: 
\begin{align} \label{Ergodicity4}
	\Big\{\omega: \Big(&{f} \big({y}_{k}(\omega), {y}_{k+1}(\omega),...\big) , {f}\big({y}_{k+1}(\omega), {y}_{k+2}(\omega),...\big), \nonumber\\
	& ... \Big) \in B^e_{\xi} \Big \} = \Big\{\omega: \big({y}_{k}(\omega), {y}_{k+1}(\omega),...\big)  \in B^e_{y} \Big \}.
\end{align}
Since $A$ is an invariant set for $(\xi_k)_{k \geq 1}$, the set $\Big\{\omega: \big({y}_1(\omega), {y}_{2}(\omega),...\big)  \in B^e_{y} \Big \}$ is also an invariant set for $(y_k)_{k \geq 1}$. Since  $(y_k)_{k \geq 1}$ is ergodic and using \eqref{Ergodicity4}, $P(A)$ is zero or one which implies that the process $(\xi_k)_{k \geq 1}$ is ergodic.

\section{Sufficient Condition for Ergodicity of Stationary Discrete-time  Gaussian Vector Processes}\label{Proof_of_Ergodocity_For_GP}
\begin{thm} Suppose $(y_k)_{k \geq 1}$ is a stationary Gaussian vector process. If  the auto- and cross-covariances vanish as the lag  $\kappa$ approaches infinity, then $(y_k)_{k \geq 1}$ is ergodic.
\end{thm}
\begin{pf}
	The proof follows the same line of 
	\cite{CramerLeadbetter} and references therein with modifications to allow stationary discrete-time Gaussian vector processes. Let us first mention a theorem  useful for the proof. 	
	\begin{thm} \label{Littlewood}   \cite{Breiman} Suppose $(\Omega, \mathcal{F}, P)$ is a probability space and $\mathcal{A}$ is a field of subsets of $\Omega$ such that the sigma-field containing $\mathcal{A}$ is $\mathcal{F}$. For any $\epsilon > 0$ and every $ C\in \mathcal{F}$ there is a set $\tilde{C} \in \mathcal{A}$ such that
		\begin{align} 
			P(C \triangle \tilde{C}) \leq \epsilon,
		\end{align}
		where $\triangle$ is symmetric difference and defined as $C \triangle \tilde{C}:=(C-\tilde{C}) \cup (\tilde{C}-C)$.
	\end{thm}
	The events in the sigma-field $\mathcal{B}_\infty$ in $\mathbb{R}^\infty$ can be approximated in probability by finite dimensional sets. If $(\mathbb{R}^\infty, \mathcal{B}_\infty, P)$ is a probability space and $C \in \mathcal{B}_\infty$, then for any $\epsilon>0$, there is a finite $n$ and an event $C_n \in \mathcal{B}_n$, where $\mathcal{B}_n$ is the $\sigma$-field in $\mathbb{R}^n$, such that 
	\begin{align} 
		P(C \triangle \tilde{C}_n) \leq \epsilon,
	\end{align}
	where $\tilde{C}_n=\{x \in \mathbb{R}^\infty: (x_1,...,x_n) \in C_n\}$ \cite{Lindgren}. 
	
	Now we turn to the proof. Let us assume that $A:=\{\omega:(y_k(\omega) , y_{k+1}(\omega),  ... ) \in B\}$ is an invariant set for  $ (y_k )_{k\geq 1}  $.  
	Define:
	\begin{align} 
		A_{\kappa}:=\{\omega:(y_{k+\kappa}(\omega),  y_{k+\kappa+1}(\omega), ... ) \in B\}, \ \ \kappa\geq 0.
	\end{align}
	It follows from stationarity that $P(A)=P(A_{\kappa})$ and since $A$ is an invariant set, the following holds:
	\begin{align} 
		P(A)=P(A_{\kappa})=P(A \cap A_{\kappa}).
	\end{align}
	Theorem \ref{Littlewood} implies that for any $\epsilon>0$, there exist a finite $n$ and $\tilde{A}_n$ such that:
	\begin{align} \label{SymSandI1}
		P(A \triangle \tilde{A}) \leq \epsilon,
	\end{align}
	where $\tilde{A}=\{\omega: (y_k(\omega),y_{k+1}(\omega),   ...,   y_{k+n}(\omega) ) \in \tilde{A}_n\}$. Using the fact that for any two random events $E_1$ and $E_2$, $|P(E_1) - P(E_2)| \leq P(E_1 \triangle E_2)$ yields that:
	\begin{align} 
		|P(A) - P(\tilde{A})| \leq \epsilon.
	\end{align}
	Now let us define
	\begin{equation} 
		\tilde{A}_{\kappa}=\{\omega: (y_{k+\kappa}(\omega) , y_{k+\kappa+1}(\omega) , ...,  y_{k+\kappa+n}(\omega) ) \in \tilde{A}_n\}.
	\end{equation}
	By stationarity, the following holds:
	\begin{align} \label{SymSandI2}
		P(A \triangle \tilde{A}_{\kappa}) \leq \epsilon.
	\end{align}
	Using \eqref{SymSandI1} and \eqref{SymSandI2}:
	\begin{align} \label{SymSandI3}
		|P(A) - P(\tilde{A}  \cap \tilde{A}_{\kappa})| &\leq P \big (A \triangle (\tilde{A}  \cap \tilde{A}_{\kappa})\big) \nonumber \\ &\leq  P(A \triangle \tilde{A} ) + P(A \triangle \tilde{A}_{\kappa}) \nonumber\\ &\leq 2\epsilon,
	\end{align}
	where the second inequality follows from the fact that for any three random events $E_1$, $E_2$ and $E_3$,  $	E_1 \triangle \big( E_2 \cap E_2) \subseteq   ( E_1 \triangle  E_{2}) \cup   ( E_1 \triangle  E_{3})$. \\
	Since the random vector process is jointly Gaussian and based on the assumption, the auto- and cross-covariances go to zero when $\kappa$ approaches infinity, it follows that:
	\begin{align} 
		\lim\limits_{\kappa \to \infty} P(\tilde{A}  \cap \tilde{A}_{\kappa}) = P(\tilde{A}) P(\tilde{A}_{\kappa}).
	\end{align}
	By stationarity of the random process $(y_k)_{k \geq 1}$ ($P(\tilde{A})=P(\tilde{A}_{\kappa})$), it results that:
	\begin{align} \label{SymSandI41} 
		\lim\limits_{\kappa \to \infty} P(\tilde{A} \cap \tilde{A}_{\kappa}) = P(\tilde{A})^2.
	\end{align}
	Equations \eqref{SymSandI2}, \eqref{SymSandI3} and  \eqref{SymSandI41} and stationarity imply that:
	\begin{align} \label{SymSandI4} 
		P(\tilde{A})=\lim\limits_{\kappa \to \infty} P(\tilde{A}_{\kappa})  =P(A),
	\end{align}
	and
	\begin{align} \label{SymSandI5}
		P(\tilde{A})^2 = \lim\limits_{\kappa \to \infty} P(\tilde{A} \cap \tilde{A}_{\kappa}) = P(A).
	\end{align}
	Using \eqref{SymSandI4} and \eqref{SymSandI5}, it follows that:
	\begin{align} 
		P(A)^2 = P(A),
	\end{align}
	which means that $P(A)$ is zero or one. Therefore, the stationary sequence $(y_k)_{k \geq 1}$ is ergodic.
\end{pf}

\end{document}